\documentclass[pdflatex,sn-mathphys,Numbered]{sn-jnl}

\usepackage{graphicx}%
\usepackage{multirow}%
\usepackage{amsmath,amssymb,amsfonts}%
\usepackage{amsthm}%
\usepackage{mathrsfs}%
\usepackage[title]{appendix}%
\usepackage{xcolor}%
\usepackage{textcomp}%
\usepackage{manyfoot}%
\usepackage{booktabs}%
\usepackage{algorithm}%
\usepackage{algorithmicx}%
\usepackage{algpseudocode}%
\usepackage{listings}%
\usepackage[font=small]{subcaption}

\numberwithin{equation}{section}
\newtheorem{theorem}{Theorem}[section]
\newtheorem{conj}{Conjecture}[section]

\theoremstyle{definition}

\theoremstyle{remark}

\newcommand{\e}[1]{{\varepsilon_#1}}
\newcommand{\w}[1]{{\omega_#1}}
\newcommand{\g}[1]{{\gamma_#1}}

\newcommand{\Cmainw}{\mathcal{C}_0}
\newcommand{\Cotherw}{\mathcal{C}_1}
\newcommand{\Cmainz}{\mathcal{C}_0'}
\newcommand{\Cotherz}{\mathcal{C}_1'}

\newcommand{\bconst}{B}
\newcommand{\circlecontour}{C}
\newcommand{\cov}{\operatorname{cov}}

\begin{document}

\title[A numerical study of two-point correlation functions of the two-periodic weighted Aztec diamond in mesoscopic limit]{A numerical study of two-point correlation functions of the two-periodic weighted Aztec diamond in mesoscopic limit}


\author*[1]{\fnm{Emily} \sur{Bain}}\email{emily\_bain@berkeley.edu}

\affil*[1]{\orgdiv{Department of Mathematics}, \orgname{University of California, Berkeley}, \orgaddress{\street{Evans Hall}, \city{Berkeley}, \postcode{94720}, \state{California}, \country{USA}}}


\abstract{In \cite{Bain_2023}, we found asymptotics of one-point correlation functions of the two-periodic weighted Aztec diamond in the mesoscopic limit, where the linear size of the ordered region is of the same order as the correlation length. In this paper, we follow up with a numerical study of two-point correlation functions of dimers separated by a mesoscopic distance.}




\maketitle

\section{Introduction}

A \textit{dimer model} is a probability distribution on the set of \textit{dimer configurations} (perfect matchings on a planar graph). It is convenient to consider only bipartite graphs, in which case a height function \cite{Thurston1990} can be defined on the faces of the dimer graph. This means there is a bijection between dimer models and random surfaces. For certain dimer models, there is a variational principle \cite{Cohn_2000} that can be used to find the limit shape of this random surface as the graph size tends to infinity \cite{kenyon2007limit, Astala2020DimerMA}.

In this paper, we look at a particular graph known as the \textit{Aztec diamond graph}, which was first studied in this context in \cite{elkies1992_1, elkies1992_2}. This is part of a square grid with boundaries at 45 degree angles to the grid. 
Specifically, we look at the \textit{two-periodic weighted Aztec diamond} \cite{chhita2014, Francesco_2014}, which is a probability measure on dimer configurations of the Aztec diamond graph with doubly-periodic weights (Figure~\ref{fig:aztec diamond diagram}).

Dimer models exhibit up to three different phases characterized by the rate of decay of correlation functions between dimers or equivalently by the variance of the height function. These three phases are known as \textit{frozen} (where correlation functions are translation invariant), \textit{disordered} (where two-point correlation functions decay quadratically) and \textit{ordered} (where two-point correlation functions decay exponentially). The two-periodic weighted Aztec diamond is one of the simplest models to exhibit all three phases. There are phase boundaries between the frozen and disordered phases, where fluctuations of correlation functions converge to the Airy process \cite{Johansson2005}, and between the disordered an ordered phases, which is an area of very active research. A formula for the entries of the inverse Kasteleyn matrix was found in \cite{chhita2014} and simplified in \cite{chhita2016domino}. Since then, other approaches have also been used to find the correlation functions \cite{Duits2021, BERGGREN2019}. Much research focuses on the Airy kernel point processes that have been observed at the boundary \cite{chhita2016domino, johansson2018, Beffara2018, Beffara2022, JohanssonMason2021}. There has also been progress on finding microscopic boundary paths \cite{Beffara2022, JohanssonMason2023}. 

In \cite{Bain_2023}, we looked at a scaling limit where the weights tend towards the uniform weighting, in which the ordered region has ``mesocopic width'' in the thermodynamic limit, and found that the one-point correlation functions are no longer described by an Airy kernel point process, but by a new process. A similar phenomenon was found numerically for the six-vertex model in \cite{Belov2022TheTC}. In this paper, we do an experimental study of two-point correlation functions in the same limit.

\subsection{Overview of paper}
In \cite{Bain_2023}, we proved asymptotic formulas for the inverse Kasteleyn matrix of the two-periodic weighted Aztec diamond, for pairs of vertices that do not grow further apart as the size of the graph $n = 4m$ tends to infinity in the mesoscopic limit. This paper follows on from \cite{Bain_2023}. We conjecture asymptotic formulas for the inverse Kasteleyn matrix for pairs of vertices that are separated by a distance of order $m^{1/2}$ as the size of the graph tends to infinity, from which we can find the asymptotic two-point correlation functions. We then run Markov chain simulations to sample a large number of tilings from the appropriate distribution, and compare the two-point correlation functions to our conjectured asymptotics.

\subsection{Overview of model and Kasteleyn method}
We briefly describe the two-periodic weighted Aztec diamond and the Kasteleyn method. For full details, see \cite{Bain_2023}.

\begin{figure}[ht]
\centering
 \includegraphics[width = 0.8\textwidth]{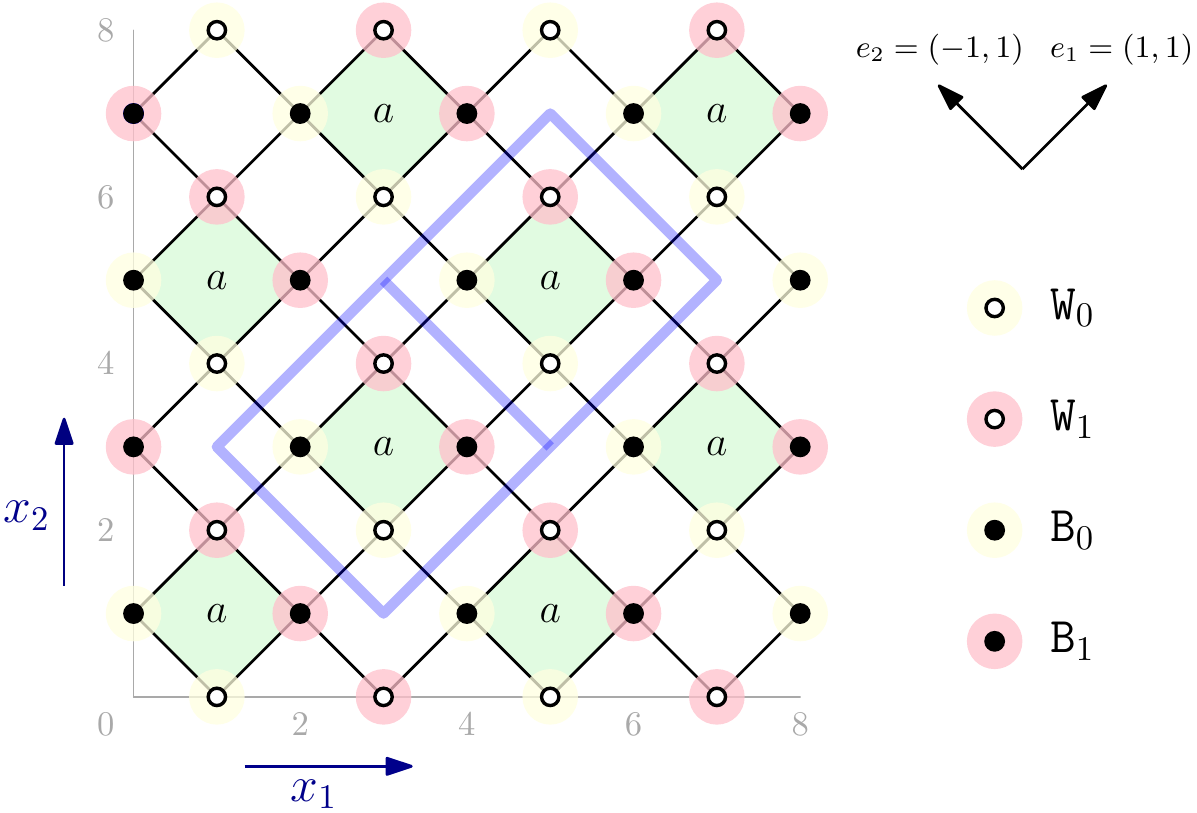}
\caption{The two periodic Aztec diamond graph of size $n=4$ showing the subgraphs $\mathtt{W}_0$, $\mathtt{W}_1$, $\mathtt{B}_0$ and $\mathtt{B}_1$. Edges surrounding a face labeled $a$ have weight $a$; other edges have weight 1. Two fundamental domains are marked in blue.}
\label{fig:aztec diamond diagram}
\end{figure}

We denote the dimer graph by $\Gamma$. We consider graphs with linear size $n = 4m$, with coordinates $(x_1, x_2)$, $x_i\in\mathbb{Z}$ for $0\leq x_i \leq 8m$, with vertices at points where $x_1 + x_2 \equiv 1 \mod 2$, as in Figure~\ref{fig:aztec diamond diagram}. We color vertices alternately black and white as shown. We denote the set of white vertices by $\mathtt{W}$ and the set of black vertices by $\mathtt{B}$ where \begin{align*}
\mathtt{W} &= \{ (x_1,x_2) \in \Gamma : x_1 \equiv 1 \mod 2,
\, x_2 \equiv 0 \mod 2,\, 0\leq x_1,x_2 \leq 2n\}\\
\mathtt{B} &= \{ (y_1,y_2) \in \Gamma : y_1 \equiv 0 \mod 2,
\, y_2 \equiv 1 \mod 2,\, 0\leq y_1,y_2 \leq 2n\}.\end{align*}
These sets of vertices are further split into two subsets each as follows. For $\e1,\e2\in \{0,1\}$, define \begin{align*}
\mathtt{W}_\e1 &= \{ (x_1,x_2) \in \mathtt{W} : x_1 + x_2 \equiv 2\e1 + 1 \mod 4\} \\
\mathtt{B}_\e2 &= \{ (y_1,y_2) \in \mathtt{B} : y_1 + y_2 \equiv 2\e2 + 1 \mod 4\}. 
\end{align*} These subsets are indicated in Figure~\ref{fig:aztec diamond diagram} in yellow and pink.

Let $e_1 = (1,1)$ and $e_2 = (-1,1)$. A $2\times 2$ fundamental domain \cite{kenyon_okounkov_sheffield_2006} embedded in the graph consists of vertices $\mathbf{w}\in \mathtt{W_0},\,\mathbf{w} + e_1 \in \mathtt{B_1}, \mathbf{w} + e_2\in \mathtt{B_0}$ and $\mathbf{w}+e_1 + e_2 \in \mathtt{W_1}$.

The edges are assigned weights $a$ and 1 as shown in Figure~\ref{fig:aztec diamond diagram}.

Let $\Omega$ denote the set of all dimer configurations on $\Gamma$. The Boltzmann measure on dimer configurations is defined as \[
\mathrm{Prob}(\mathcal{D}) = \frac{\prod_{e\in \mathcal{D}}w(e)}{\sum_{\mathcal{D} \in \Omega }\prod_{e\in \mathcal{D}}w(e)}
\] for $\mathcal{D} \in \Omega$.

Let $e_1, e_2 $ be edges of $\Gamma$. Then the two-point correlation function is defined as
\[
\rho(e_1,e_2) =  \sum_{\mathcal{D}\in \Omega} \mathrm{Prob}(\mathcal{D}) \sigma_{e_1}(\mathcal{D}) \sigma_{e_2}(\mathcal{D})
\] where \[\sigma_{e}(\mathcal{D}) = \begin{cases} 1 &\text{if } e\in \mathcal{D} \\
0 &\text{otherwise}. \end{cases}\]

We use the Kasteleyn method to compute two-point correlation functions. Let $K_a$ denote the Kasteleyn matrix of the two-periodic weighted Aztec diamond. For $x\in\mathtt{W}$ and $y\in\mathtt{B}$, we have \[
K_a(y,x) = \begin{cases}
a(1-\varepsilon) + \varepsilon & y = x + e_1,\, x\in W_\varepsilon \\
(1-\varepsilon) + a \varepsilon & y = x - e_1,\, x\in W_\varepsilon \\
i(a(1-\varepsilon) + \varepsilon) & y = x + e_2,\, x\in W_\varepsilon \\
i((1-\varepsilon) + a\varepsilon) & y = x - e_2,\, x\in W_\varepsilon \\
0 & \text{if } (x,y) \text{ is not an edge}
\end{cases}
\]

For edges $e_1 = (\mathbf{w}_1, \mathbf{b}_1), e_2 = (\mathbf{w}_2, \mathbf{b}_2)$ of $\Gamma$, the two-point correlation is given by \cite{Kenyon_1997} \begin{equation}\label{eq:rho_twopoint}
\rho(e_1, e_2) = \left(\prod_{i=1}^{2} K_a(\mathbf{b}_i,\mathbf{w}_i)\right) \det (K_a^{-1}(\mathbf{w}_i, \mathbf{b}_j))_{1 \leq i,j \leq 2}.
\end{equation}

\section{Asymptotic two-point correlation functions}
\subsection{Conjecture}
We look at the inverse Kasteleyn matrix $K_a^{-1}(x,y)$ for $x = (x_1,x_2) \in \mathtt{W}_\e1$ and $y= (y_1,y_2) \in \mathtt{B}_\e2$ that are a Euclidean distance of order $m^{1/2}$ both from the center $(4m, 4m)$ of the Aztec diamond and from each other,  in the limit where the weight $a$ is given by $a=1-Bm^{-1/2}$ for some constant $B>0$. We compute asymptotics for $x$ and $y$ near the diagonal in the third quadrant. We define the asymptotic coordinates $\alpha_x,\, \alpha_y < 0$ as follows. \begin{align}\begin{split}\label{eq:asymptoticcoord}
x_1 &= [4m + 2m^{1/2}\alpha_x \bconst] + \overline{x_1} \\ 
x_2 &= [4m + 2m^{1/2}\alpha_x \bconst] + \overline{x_2} \\ 
y_1 &= [4m + 2m^{1/2}\alpha_y \bconst] + \overline{y_1} \\ 
y_2 &= [4m + 2m^{1/2}\alpha_y \bconst] + \overline{y_2},
\end{split}
\end{align} where the integral parts $\overline{x}_i, \overline{y}_i \in \mathbb{Z}$ do not grow with $m$. In \cite{Bain_2023} we considered the case where $\alpha_x = \alpha_y$. Here we consider the case $\alpha_x \neq \alpha_y$.

As in \cite{Bain_2023}, we define the matrix $\zeta$ to have entries \begin{equation}\label{eq:zeta}
\zeta(x,y)=(-1)^{(y_2-x_1)/2}.
\end{equation} We also define the matrix $\Sigma$ by \begin{equation}\label{eq:Sigmaxy}
\Sigma(x,y) = \begin{cases}
1 & y = x + (2k+1)e_1 + 2le_2, \text{ some } k,l\in \mathbb{Z}\\
i & y = x + 2ke_1 + (2l+1)e_2, \text{ some } k,l\in \mathbb{Z}
\end{cases}
\end{equation} 

Let $\eta = \eta(\alpha)$ be defined to be the unique complex number with non-negative real part and non-negative imaginary part that satisfies \begin{equation}
\frac{1}{\sqrt{1/2 - 2 i \eta}} + \frac{1}{\sqrt{1/2+ 2i \eta}} = -2/\alpha.\label{eq:eta}\end{equation} and let $\eta'$ be the unique complex number that satisfies \begin{equation}
\frac{1}{\sqrt{1/2 - 2 i \eta'}}- \frac{1}{\sqrt{1/2+ 2i \eta'}} = 2/\alpha.\label{eq:eta'}\end{equation} For all $\alpha < 0$ we have $\eta' \in i(0, 1/4)$.

Let \begin{equation}\label{eq:fpm}
f^\pm(w) = \sqrt{1/2-2iw} \pm \sqrt{1/2+2iw}.
\end{equation}and for $j,k \in \{0,1\}$ and $\e1,\e2 \in \{0,1\}$, let
\begin{multline}
A^{j,k}_{\e1,\e2}(w,z) =  \frac{-(-1)^{\e1+\e2}}{\sqrt{1/2-2iw}\sqrt{1/2+2iw}\sqrt{1/2-2iz}\sqrt{1/2+2iz}}\Bigg(2 i(w-z) \\
+ (-1)^{\e1+\e2}\left(\sqrt{1/2-2iw} +  (-1)^{j}\sqrt{1/2-2iz}\right)\left((-1)^{k}\sqrt{1/2+2iw} + \sqrt{1/2+2iz}\right) \\
+\Big((-1)^\e1\sqrt{1/2-2iw} + (-1)^{\e2+k}\sqrt{1/2+2iw} 
+ (-1)^\e2\sqrt{1/2+2iz} + (-1)^{\e1+j}\sqrt{1/2-2iz}\Big)\\
\times \left(\sqrt{1/2-2iw}\sqrt{1/2+2iz} + (-1)^{j+k}\sqrt{1/2+2iw}\sqrt{1/2-2iz}\right)\Bigg).\label{eq:Aall}
\end{multline} 
We define the following double integrals, where $x = (x_1,x_2) \in \mathtt{W}_\e1$ and $y= (y_1,y_2) \in \mathtt{B}_\e2$ and $\alpha_x,\,\alpha_y$ are as in Equation~\ref{eq:asymptoticcoord}. The integrals defined in \cite{Bain_2023} are the same as these, but for $\alpha_x = \alpha_y = \alpha$.
\begin{align}\begin{split}
I_1(\alpha_x, \alpha_y, \e1, \e2) &=  \int_{\Cmainw} dw\int_{\Cmainz}dz \,\frac{A_{\e1,\e2}^{0,0}(w,z)}{i(z-w)} \exp(\bconst^2(-2i(w-z)+ \alpha_x f^-(w)  - \alpha_y f^-(z))),\\
I_2(\alpha_x, \alpha_y, \e1, \e2) &=  \int_{\Cmainw} dw\int_{\Cotherz}dz \,\frac{A_{\e1,\e2}^{1,0}(w,z)}{i(z-w)} \exp(\bconst^2(-2i(w-z)+ \alpha_x f^-(w)  +  \alpha_y f^+(z))) ,\\
I_3(\alpha_x, \alpha_y, \e1, \e2) &=  \int_{\Cotherw} dw\int_{\Cmainz}dz \,\frac{A_{\e1,\e2}^{0,1}(w,z)}{i(z-w)} \exp(\bconst^2(-2i(w-z)+ \alpha_x f^+(w)  -  \alpha_y f^-(z))) ,\\
I_4(\alpha_x, \alpha_y, \e1, \e2) &=  \int_{\Cotherw} dw\int_{\Cotherz}dz \,\frac{A_{\e1,\e2}^{1,1}(w,z)}{i(z-w)} \exp(\bconst^2(-2i(w-z)+ \alpha_x f^+(w) + \alpha_y  f^+(z))),\label{eq:I1to4}\end{split}
\end{align} where the functions $A_{\e1,\e2}^{j,k}(w,z)$ are defined in Equation~\ref{eq:Aall} and the contours are defined below. When either $\alpha_x < -1/\sqrt{2}$ or $\alpha_y < -1/\sqrt{2}$, we also define the single integral
 \begin{equation}
I_0(\alpha_x, \alpha_y, a, \e1, \e2) =  \int_{\gamma}\frac{1 + (-1)^\e2\sqrt{1/2-2iw} + (-1)^\e1\sqrt{1/2+2iw}}{\sqrt{1/2-2iw}\sqrt{1/2+2iw}}dw.
\label{eq:I0}\end{equation}

The contours $\Cmainw$, $\Cmainz$, $\Cotherw$, $\Cotherz$ and $\gamma$ are defined as follows.

Recall that along a steepest descent contour of a holomorphic function (contour where the real part decreases most rapidly), its imaginary part is constant. A function has a saddle point when its second derivative is 0. The function $-2iw + \alpha f^-(w)$ has saddle points at $w = \pm \eta$ and the function $-2iw + \alpha f^+(w)$ has a saddle point at $w = -\eta'$.

For $-1/\sqrt{2}<\alpha_x < 0$, let $\Cmainw$ be the steepest descent contour for $-2iw + \alpha_x f^-(w)$ that is contained in the negative half plane and passes through the saddle point $w = -\eta$.

For $\alpha_x = -1/\sqrt{2}$ let $\Cmainw$ be the steepest descent contour for $-2iw + \alpha_x f^-(w)$ that passes through the saddle point $w = 0$ and enters the negative half plane at angles of $-\pi/6$ and $-5\pi/6$.

For $\alpha_x < -1/\sqrt{2}$, let $\Cmainw$ consist of the steepest descent contour for $-2iw + \alpha_x f^-(w)$ that starts from the branch cut $i(1/4,\infty)$, passes through the saddle point $w = -\eta$ and goes to infinity in the third quadrant; the reflection in the imaginary axis of this contour; and a contour that goes around the branch cut $i(1/4, \infty)$.

For $-1/\sqrt{2}<\alpha_y < 0$, let $\Cmainz$ be the steepest descent contour for $2iz - \alpha_y f^-(z)$ that is contained in the positive half plane and passes through the saddle point $z = \eta$.

For $\alpha_y = -1/\sqrt{2}$ let $\Cmainz$ be the steepest descent contour for $2iz - \alpha_y f^-(z)$ that passes through the saddle point $z = 0$ and enters the positive half plane at angles of $\pi/6$ and $5\pi/6$.

For $\alpha_y < -1/\sqrt{2}$, let $\Cmainz$ consist of the steepest descent contour for $2iz - \alpha_y f^-(z)$ that starts from the branch cut $i(-\infty, -1/4)$, passes through the saddle point $z = \eta$ and goes to infinity in the second quadrant; the reflection in the imaginary axis of this contour; and a contour that goes around the branch cut $i(-\infty, -1/4)$. 

Let $\Cotherw$ be the steepest descent contour for $ -2iw + \alpha_x f^+(w)$. This passes through $w = -\eta'$ and goes to infinity in the negative half plane.

Let $\Cotherz$ be the steepest descent contour for $ 2iz + \alpha_y f^+(z)$. This passes through $z = \eta'$ and goes to infinity in the positive half plane.

Note that for $\alpha_x = \alpha_y$, $\Cmainz$ is the reflection of $\Cmainw$ in the real axis, and $\Cotherz$ is the reflection of $\Cotherw$ in the real axis.

 When either $\alpha_x < -1/\sqrt{2}$ or $\alpha_y < -1/\sqrt{2}$, let the intersection points of $\Cmainw$ and $\Cmainz$ be denoted $\pm \mu$ with $\mathrm{Re}(\mu) > 0$. Let $\gamma$ be the contour composed of straight lines from $-\mu$ to $-\mathrm{Re}(\mu)$ to $\mathrm{Re}(\mu)$ to $\mu$.

In the limit as $m$ tends to infinity with $a = 1-\bconst m^{-1/2}$ we make the following conjecture for the entries of the inverse Kasteleyn matrix $K_a^{-1}$ when $\alpha_x \neq \alpha_y$.

\begin{conj}\label{conj:mainresult}
Take $x \in \mathtt{W}_\e1$ and $y\in \mathtt{B}_\e2$ with $\e1,\e2\in \{0,1\}$. Let $K_0(\alpha)$ and $K_1(\alpha)$ denote modified Bessel functions of the second kind. Recall the definitions of $\zeta(x,y)$ and $\Sigma(x,y)$ from Equations~\ref{eq:zeta} and \ref{eq:Sigmaxy} respectively. Let $\alpha_x,\, \alpha_y$ be the asymptotic coordinates as in Equation~\ref{eq:asymptoticcoord}. For $-1/\sqrt{2} \leq \alpha_x,\, \alpha_y < 0$ we have
\begin{multline}\label{eq:mainresult1}
K_a^{-1}(x,y) = \bconst m^{-1/2} \frac{\zeta(x,y)}{\Sigma(x,y)}\bigg(\bigg(-\frac{1}{2}K_0(\sqrt{2}\bconst^2|\alpha_y - \alpha_x|) \\+ (\e2-\e1) \operatorname{sgn}(\alpha_y-\alpha_x)\frac{1}{\sqrt{2}\pi}K_1(\sqrt{2\pi}\bconst^2|\alpha_y - \alpha_x|)\bigg) + \psi(\alpha_x, \alpha_y,\e1,\e2)\bigg) + o(m^{-1/2})
\end{multline} and for $\alpha_x < -1/\sqrt{2}$ or $\alpha_x < -1/\sqrt{2}$ we have
\begin{multline}\label{eq:mainresult2}
K_a^{-1}(x,y) = \bconst m^{-1/2} \frac{\zeta(x,y)}{\Sigma(x,y)}\bigg(\bigg(-\frac{1}{2\pi}K_0(\sqrt{2}\bconst^2|\alpha_y - \alpha_x|)\\
 + (\e2-\e1) \operatorname{sgn}(\alpha_y-\alpha_x)\frac{1}{\sqrt{2}\pi}K_1(\sqrt{2}\bconst^2|\alpha_y - \alpha_x|)\bigg) \\ + \frac{I_0(\alpha_x, \alpha_y,  \e1, \e2)}{4\pi}
+ \psi(\alpha_x, \alpha_y, \e1,\e2)\bigg)+o(m^{-1/2})
\end{multline} where \begin{multline}\label{eq:psixy} \psi(\alpha_x, \alpha_y, \e1, \e2) = \frac{1}{32 \pi^2}(I_1(\alpha_x, \alpha_y,  \e1, \e2) - I_2(\alpha_x, \alpha_y,  \e1, \e2)\\ - I_3(\alpha_x, \alpha_y,  \e1, \e2) + I_4(\alpha_x, \alpha_y,  \e1, \e2))
\end{multline} and the integrals $I_0(\alpha_x, \alpha_y,  \e1, \e2)$, $I_1(\alpha_x, \alpha_y, \e1, \e2)$, $I_2(\alpha_x, \alpha_y, \e1, \e2)$, $I_3(\alpha_x, \alpha_y, \e1, \e2)$ and $I_4(\alpha_x, \alpha_y, \e1, \e2)$ are defined above in Equations~\ref{eq:I1to4} and \ref{eq:I0}.
\end{conj}

The Bessel functions indicate a sine-Gordon field. These terms were found in \cite{Mason2022}, where the author proved that the two-point correlation function of the height field of the two-periodic weighted dimer model on the plane at the ordered-disordered boundary converges to the two-point correlation function of the sine-Gordon field as the weights tend to uniform. The other terms are specific to our scaling limit in which the size of the Aztec diamond graph tends to infinity as the weights tend to uniform. This suggests a new scaling regime.

\subsection{Sketch proof}
We provide a sketch proof of this conjecture.

We start with the following theorem from \cite{Bain_2023}, which follows from \cite{chhita2016domino} and \cite{chhita2014}.

\begin{theorem}[\cite{Bain_2023}]\label{theorem:Ka1} For $n = 4m$ and $0 <a < 1$, take $x = (x_1,x_2) \in \mathtt{W}_\e1,\, y = (y_1,y_2) \in \mathtt{B}_\e2$ with $\e1,\e2 \in \{0,1\}$. Then the entries of the inverse Kasteleyn matrix $K_a^{-1}$ are given by
\begin{multline}
K_a^{-1}((x_1,x_2),(y_1,y_2)) = \mathbb{K}_{a,0,0}^{-1}((x_1,x_2),(y_1,y_2)) - \Big(\mathcal{I}^{0,0}_{\e1,\e2} (a, x_1, x_2, y_1, y_2) \\
-\mathcal{I}^{1,0}_{\e1,\e2}(a, x_1, x_2, y_1, y_2) - \mathcal{I}^{0,1}_{\e1,\e2} (a, x_1, x_2, y_1, y_2) + \mathcal{I}^{1,1}_{\e1,\e2} (a, x_1, x_2, y_1, y_2)\Big)\label{eq:Ka1inverse}
\end{multline}
where $\mathbb{K}_{a,0,0}^{-1}((x_1,x_2),(y_1,y_2))$ is defined in Equation~\ref{eq:translationinvariantoriginal} below and $\mathcal{I}^{j,k}_{\e1,\e2} (a, x_1, x_2, y_1, y_2)$ is defined in Appendix~\ref{sec:mathcalI}.
\end{theorem} 

The following formula for $\mathbb{K}_{a,0,0}^{-1}((x_1,x_2),(y_1,y_2))$ follows from \cite{kenyon_okounkov_sheffield_2006}. Let $\circlecontour_1$ denote the unit cicle traversed in a counter-clockwise direction. Then for $\mathbf{w} = (w_1,w_2) \in \mathtt{W}_\e1$ and $\mathbf{b} = (b_1,b_2) \in \mathtt{B}_\e2$ in the same fundamental domain, where $\e1,\e2\in \{0,1\}$, and $u, v \in \mathbb{Z}$ we have \begin{equation}
\label{eq:translationinvariantoriginal}
\mathbb{K}_{a,0,0}^{-1}(\mathbf{w}, \mathbf{b} + 2ue_1 + 2ve_2) = \frac{1}{(2\pi i)^2} \int_{\circlecontour_1}\frac{dz}{z}\int_{\circlecontour_1}\frac{dw}{w} (\mathcal{K}_a(z, w)^{-1})_{\e1\e2} z^u w^v 
\end{equation}
where \begin{equation*}
P_a(z,w) = -2 -2a^2 - aw^{-1} - aw - az^{-1} - az
\end{equation*} and \begin{equation*}
\mathcal{K}_a(z,w)^{-1} = \frac{1}{P_a(z,w)}\begin{pmatrix}
i(a+w) & -(a+z) \\
-(a+z^{-1}) & i(a+w^{-1}).
\end{pmatrix}
 \end{equation*}

We conjecture the following asymptotics for the integrals $\mathcal{I}^{j,k}_{\e1,\e2}(a, x_1, x_2, y_1, y_2)$. The derivation of these formulas is essentially the same as in \cite{Bain_2023}, but we do not provide rigorous error bounds.

\begin{conj}\label{conj:Iasym}
 For $n = 4m$ and $0 <a < 1$, take $x = (x_1,x_2) \in \mathtt{W}_\e1,\, y = (y_1,y_2) \in \mathtt{B}_\e2$ with $\e1,\e2 \in \{0,1\}$. For $-1/\sqrt{2} \leq \alpha_x, \alpha_y < 0$,
  \begin{equation}
 \mathcal{I}^{0,0}_{\e1,\e2} (a, x_1, x_2, y_1, y_2) =  \frac{\zeta(x,y) \bconst m^{-1/2}}{8(2\pi i)^2\Sigma(x,y)}\int_{\Cmainw}dw \int_{\Cmainz}dz \frac{A^{0,0}_{\e1,\e2}(w,z)}{i(z-w)} e^{g_{0,0}(w,z)} + O(m^{-1}),
 \end{equation} and when $\alpha_x < -1/\sqrt{2}$ or $\alpha_y < -1/\sqrt{2}$,
 \begin{multline}
 \mathcal{I}^{0,0}_{\e1,\e2} (a, x_1, x_2, y_1, y_2) =  \frac{\zeta(x,y) \bconst m^{-1/2}}{8(2\pi i)^2\Sigma(x,y)}\\
\times\Bigg( \int_{\Cmainw}dw \int_{\Cmainz}dz \frac{A^{0,0}_{\e1,\e2}(w,z)}{i(z-w)} e^{g_{0,0}(w,z)} - 2\pi  \int_\gamma A^{0,0}_{\e1,\e2}(w,w) dw\Bigg) + O(m^{-1}).
 \end{multline} For $(j,k) \neq (0,0)$ for any $\alpha < 0$, 
 \begin{equation}
 \mathcal{I}^{j,k}_{\e1,\e2} (a, x_1, x_2, y_1, y_2) =  \frac{\zeta(x,y) \bconst m^{-1/2}}{8(2\pi i)^2\Sigma(x,y)} \int_{\mathcal{C}_j}dw \int_{\mathcal{C}_k'}dz \frac{A^{j,k}_{\e1,\e2}(w,z)}{i(z-w)} e^{g_{j,k}(w,z)} + O(m^{-1})
 \end{equation}
 where we have \begin{align}
\begin{split}
g_{0,0}(w,z) &= \bconst^2(-2i(w-z) + \alpha_x f^-(w) - \alpha_y f^-(z)) \\
g_{1,0}(w,z) &= \bconst^2(-2i(w-z) + \alpha_x f^-(w) + \alpha_y f^+(z)) \\
g_{0,1}(w,z) &= \bconst^2(-2i(w-z) + \alpha_x f^+(w) - \alpha_y f^-(z)) \\
g_{1,1}(w,z) &= \bconst^2(-2i(w-z) + \alpha_x f^+(w) + \alpha_y f^+(z))
\end{split}\label{eq:gij}
\end{align}
\end{conj}

The asymptotics of $\mathbb{K}_{a,0,0}^{-1}((x_1,x_2),(y_1,y_2))$ follow from \cite{Mason2022}.

\begin{theorem}\label{thm:Kasym}
For $x \in \mathtt{W}_\e1$ and $y \in \mathtt{B}_\e2$ with $\e1,\e2\in \{0,1\}$ and $\alpha_x,\,\alpha_y < 0$ with $\alpha_x \neq \alpha_y$, we have \begin{multline}\label{eq:Kasym}
\mathbb{K}_{a,0,0}^{-1}(x, y) = \bconst m^{-1/2} \frac{\zeta(x,y)}{\Sigma(x,y)} \bigg(-\frac{1}{2\pi}K_0(\sqrt{2}\bconst^2|\alpha_y - \alpha_x|) \\
+ (\e2-\e1) \operatorname{sgn}(\alpha_y-\alpha_x)\frac{1}{\sqrt{2}\pi}K_1(\sqrt{2}\bconst^2|\alpha_y - \alpha_x|)\bigg) + o(m^{-1/2})
\end{multline}
\begin{proof}
This follows from translation invariance of $\mathbb{K}^{-1}(x,y)$ and Theorem~2 of \cite{Mason2022}.
\end{proof}

\end{theorem}
Putting together Conjecture~\ref{conj:Iasym} and Theorem~\ref{thm:Kasym} we obtain Conjecture~\ref{conj:mainresult}.

\section{Numerical comparison of $\mathbb{K}_{a,0,0}^{-1}(x, y)$ and asymptotics}\label{sec:numericalK}

For the numerical parts of the paper we will take $\bconst = 1$.

\begin{figure}
\centering
\begin{subfigure}[t]{0.32\textwidth}
\centering
\includegraphics[width=\textwidth]{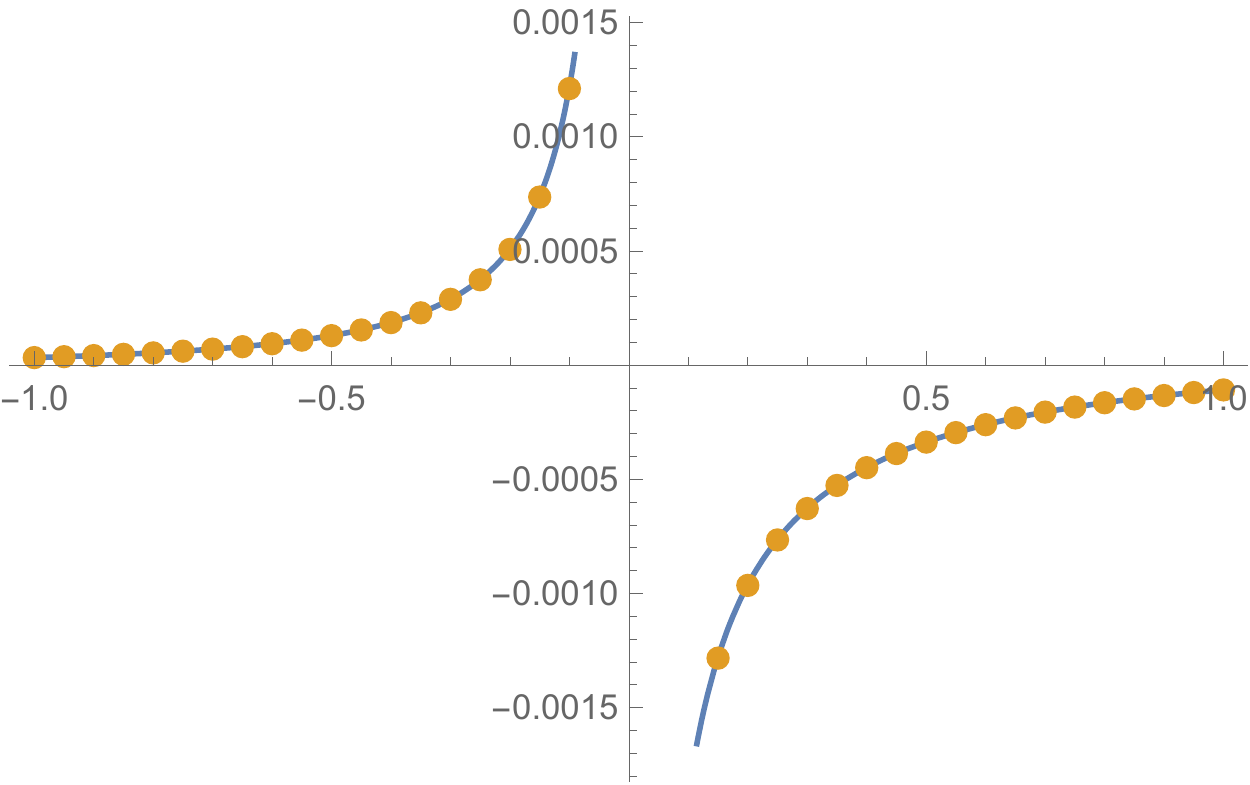}
\caption{$a = 0.999$, $v = -2$}
\end{subfigure}
\begin{subfigure}[t]{0.32\textwidth}
\centering
\includegraphics[width=\textwidth]{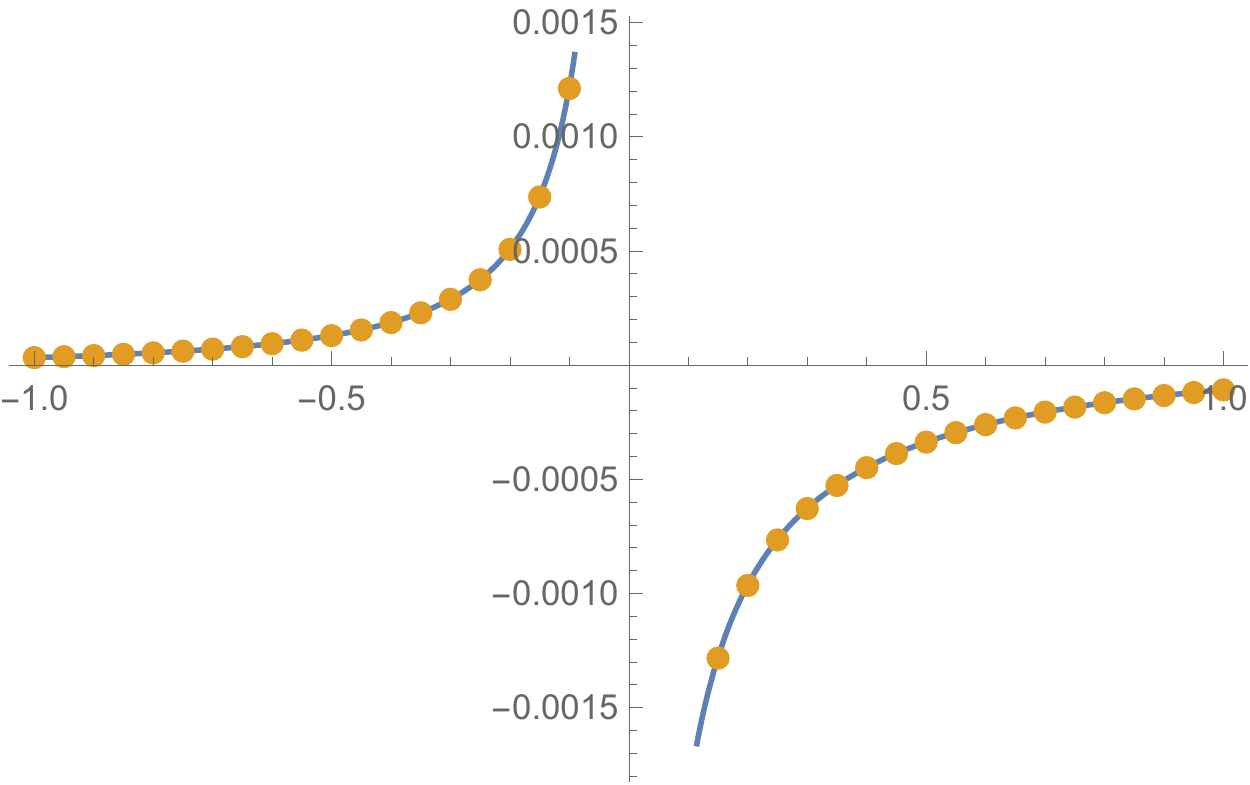}
\caption{$a = 0.999$, $v = -1$}
\end{subfigure}
\begin{subfigure}[t]{0.32\textwidth}
\centering
\includegraphics[width=\textwidth]{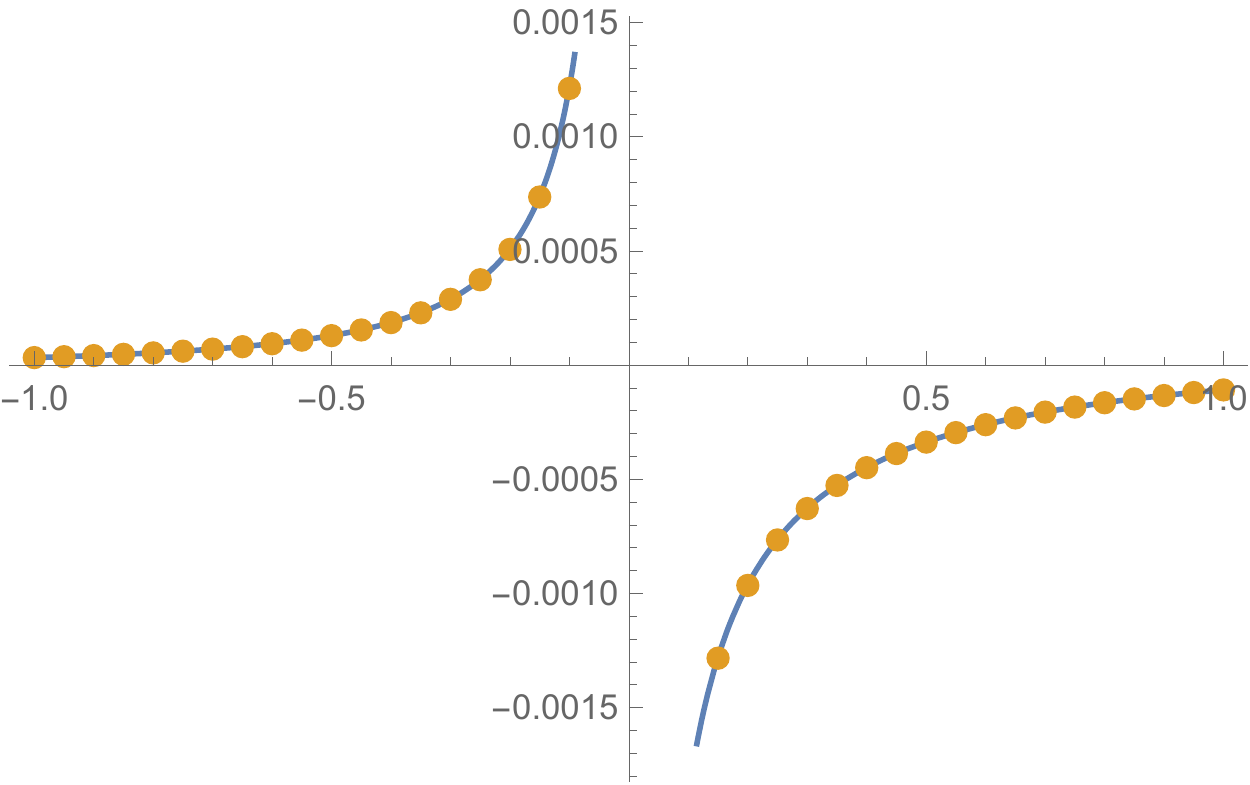}
\caption{$a = 0.999$, $v = 0$}
\end{subfigure}\\
\begin{subfigure}[t]{0.32\textwidth}
\centering
\includegraphics[width=\textwidth]{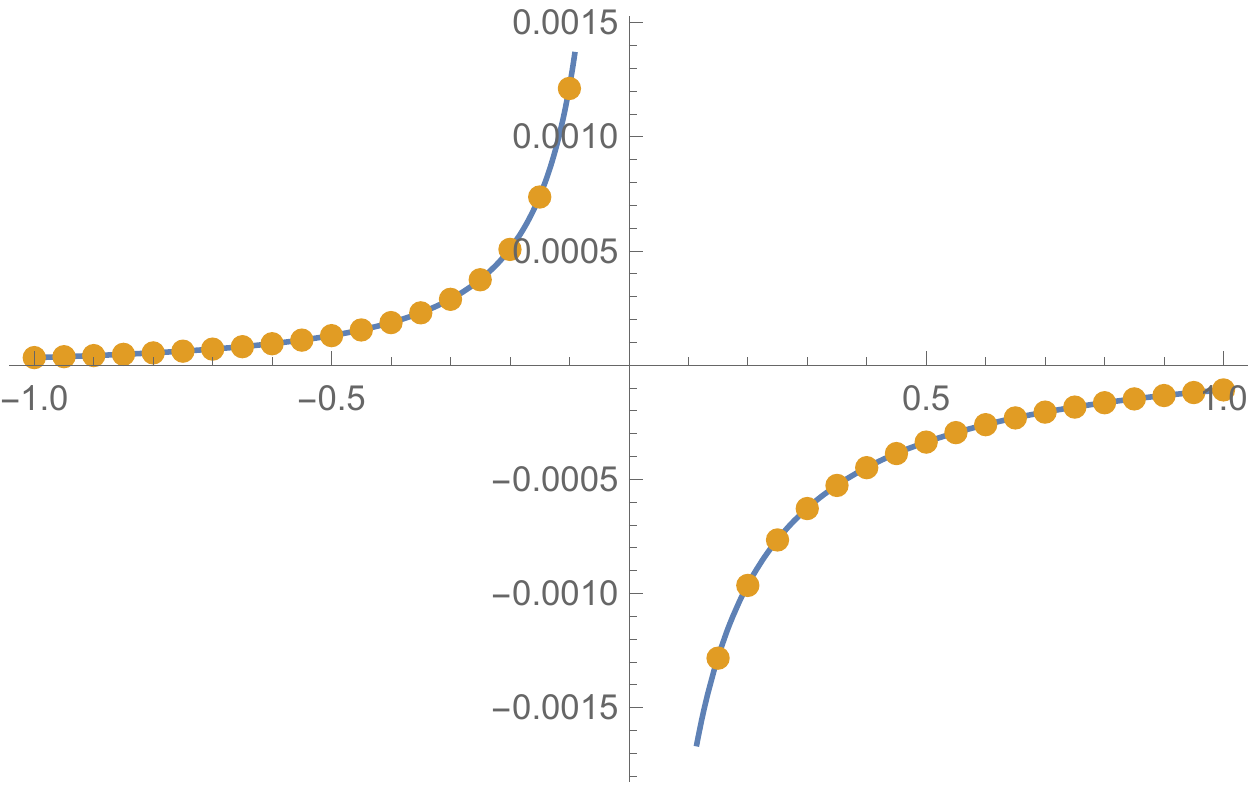}
\caption{$a = 0.999$, $v = 1$}
\end{subfigure}
\begin{subfigure}[t]{0.32\textwidth}
\centering
\includegraphics[width=\textwidth]{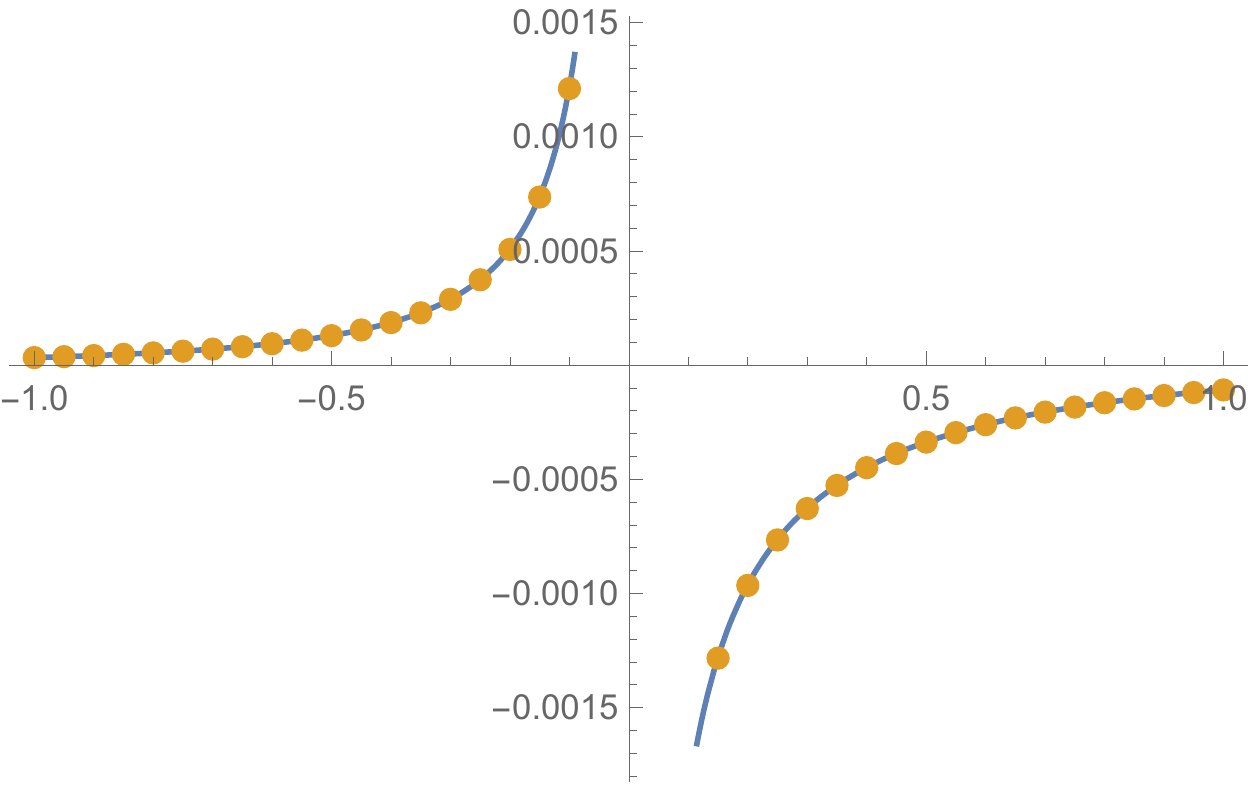}
\caption{$a = 0.999$, $v = 2$}
\end{subfigure}
\\
\begin{subfigure}[t]{0.32\textwidth}
\centering
\includegraphics[width=\textwidth]{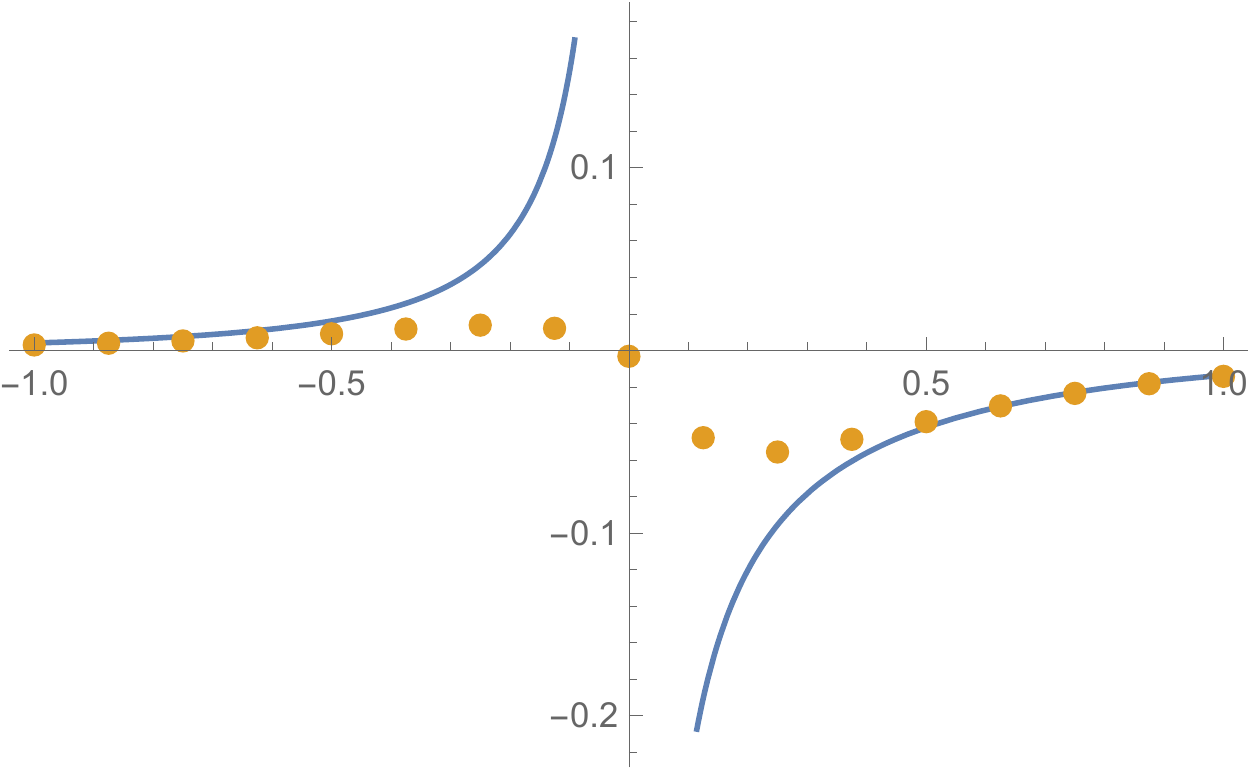}
\caption{$a = 0.875$, $v = -2$}
\end{subfigure}
\begin{subfigure}[t]{0.32\textwidth}
\centering
\includegraphics[width=\textwidth]{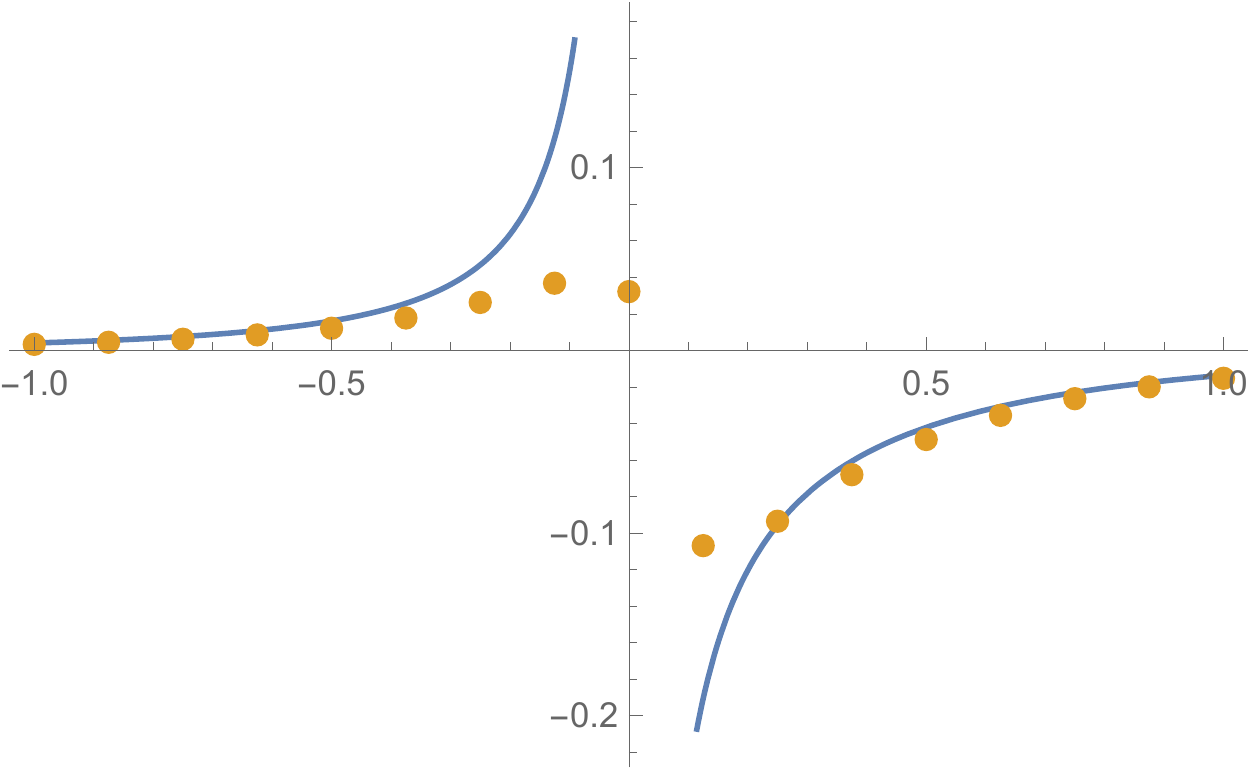}
\caption{$a = 0.875$, $v = -1$}
\end{subfigure}
\begin{subfigure}[t]{0.32\textwidth}
\centering
\includegraphics[width=\textwidth]{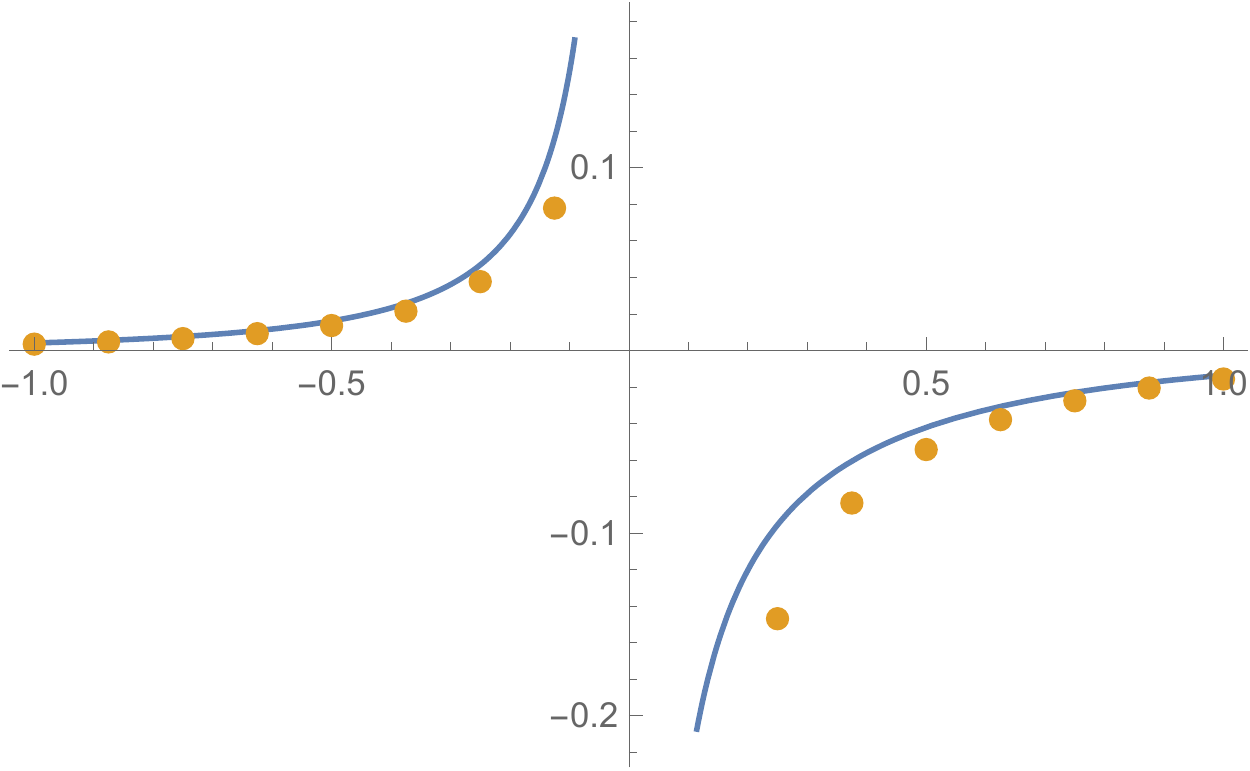}
\caption{$a = 0.875$, $v = 0$}\label{fig:k01v0}
\end{subfigure}\\
\begin{subfigure}[t]{0.32\textwidth}
\centering
\includegraphics[width=\textwidth]{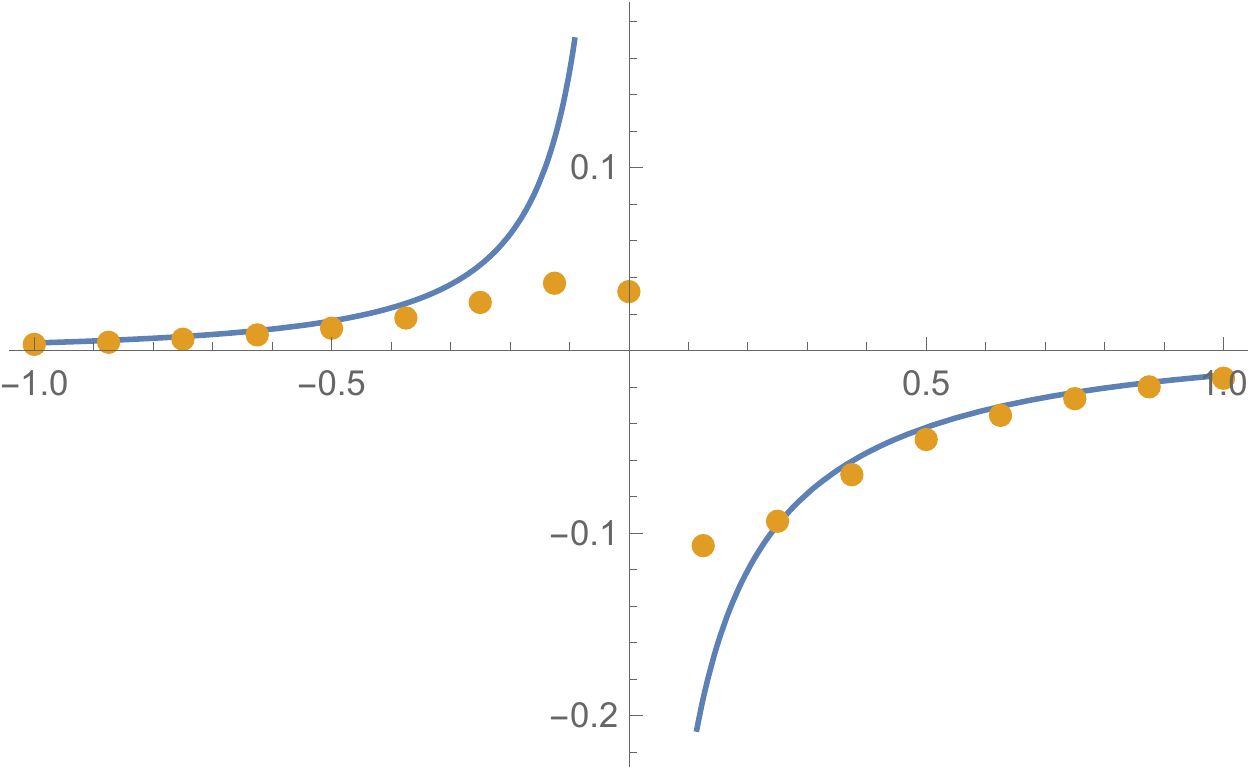}
\caption{$a = 0.875$, $v = 1$}
\end{subfigure}
\begin{subfigure}[t]{0.32\textwidth}
\centering
\includegraphics[width=\textwidth]{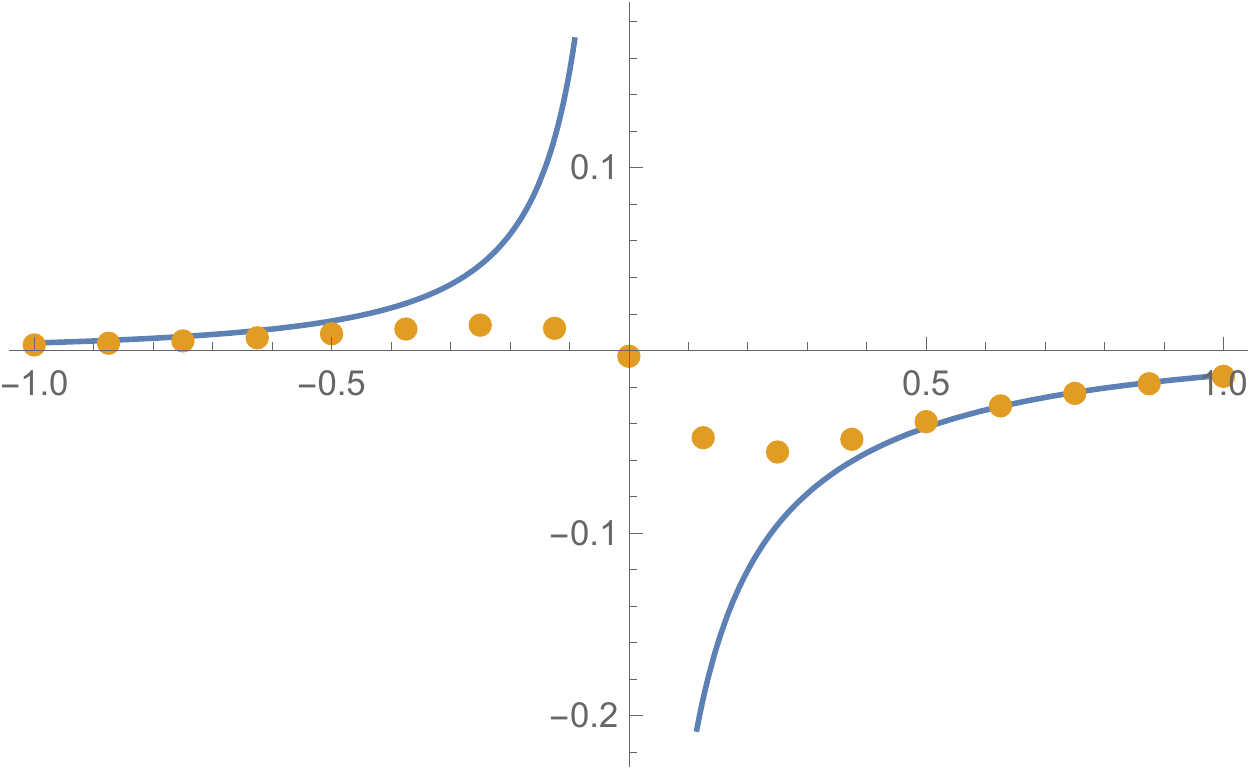}
\caption{$a = 0.875$, $v = 2$}
\end{subfigure}
\caption{The blue line shows $(1-a) (-1)^{u+v} (-K_0(\sqrt{2}\bconst^2\alpha)/2 + \operatorname{sgn}(\alpha)K_1(\sqrt{2}\bconst^2\alpha)/\sqrt{2})/\pi$ and the orange points show $(-1)^{u+v}\mathbb{K}_{a,0,0}^{-1}(\mathbf{w}_0, \mathbf{b}_1 + 2ue_1 + 2ve_2)$, both plotted against $\alpha$, where $\alpha = (1-a) u$, for various values of $a$ and $v$.}\label{fig:k01numeric}
\end{figure}

\begin{figure}
\centering
\begin{subfigure}[t]{0.32\textwidth}
\centering
\includegraphics[width=\textwidth]{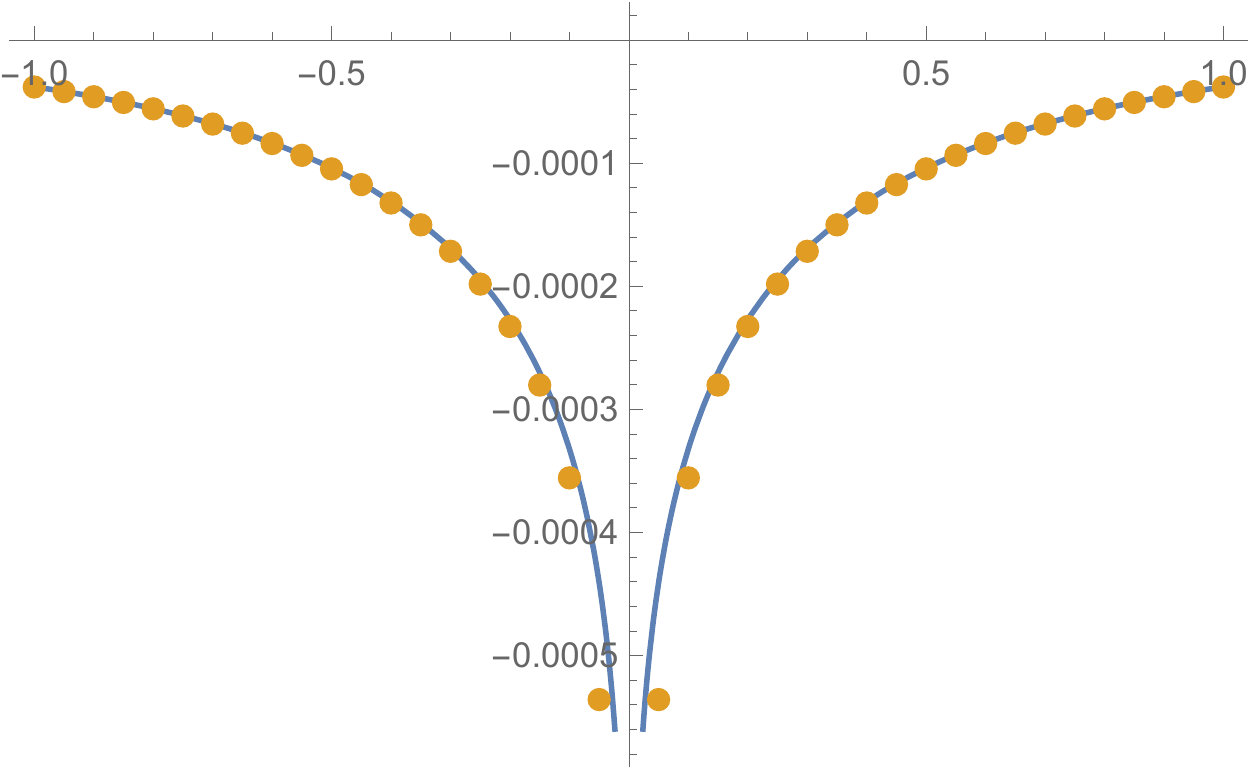}
\caption{$a = 0.999$, $v = -2$}
\end{subfigure}
\begin{subfigure}[t]{0.32\textwidth}
\centering
\includegraphics[width=\textwidth]{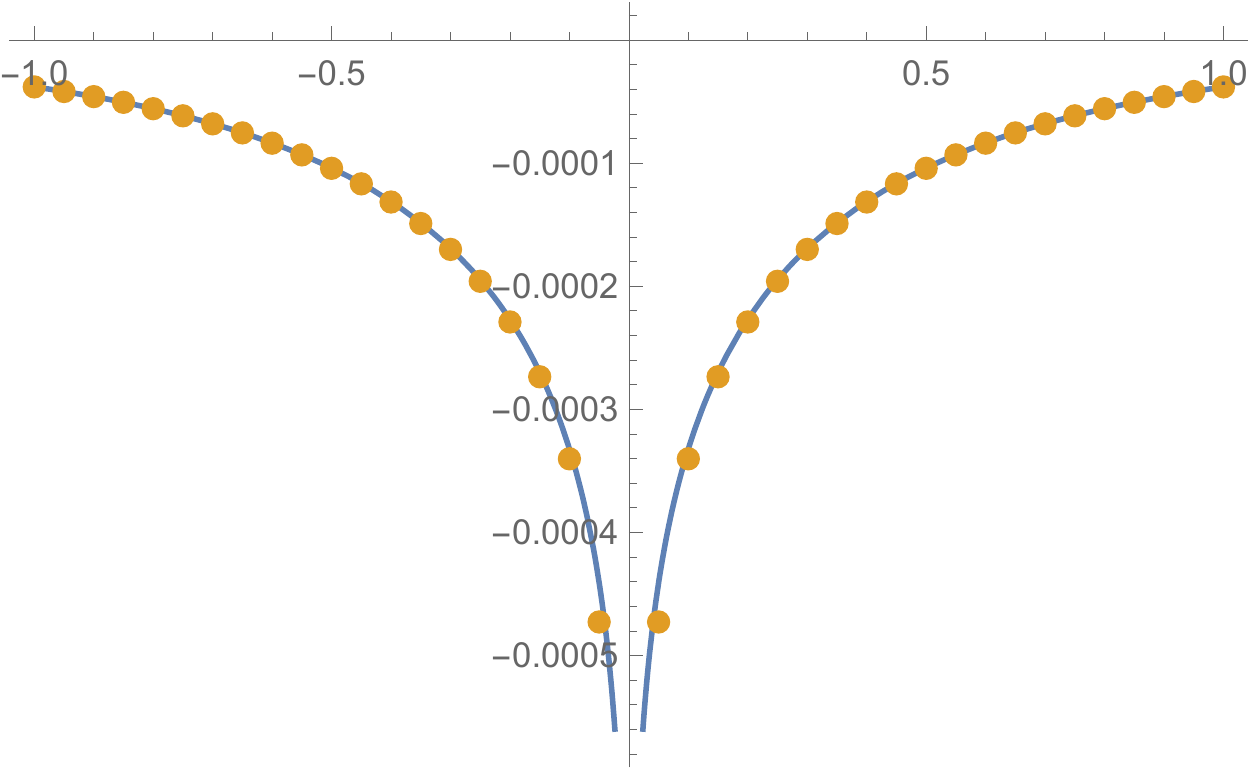}
\caption{$a = 0.999$, $v = -1$}
\end{subfigure}
\begin{subfigure}[t]{0.32\textwidth}
\centering
\includegraphics[width=\textwidth]{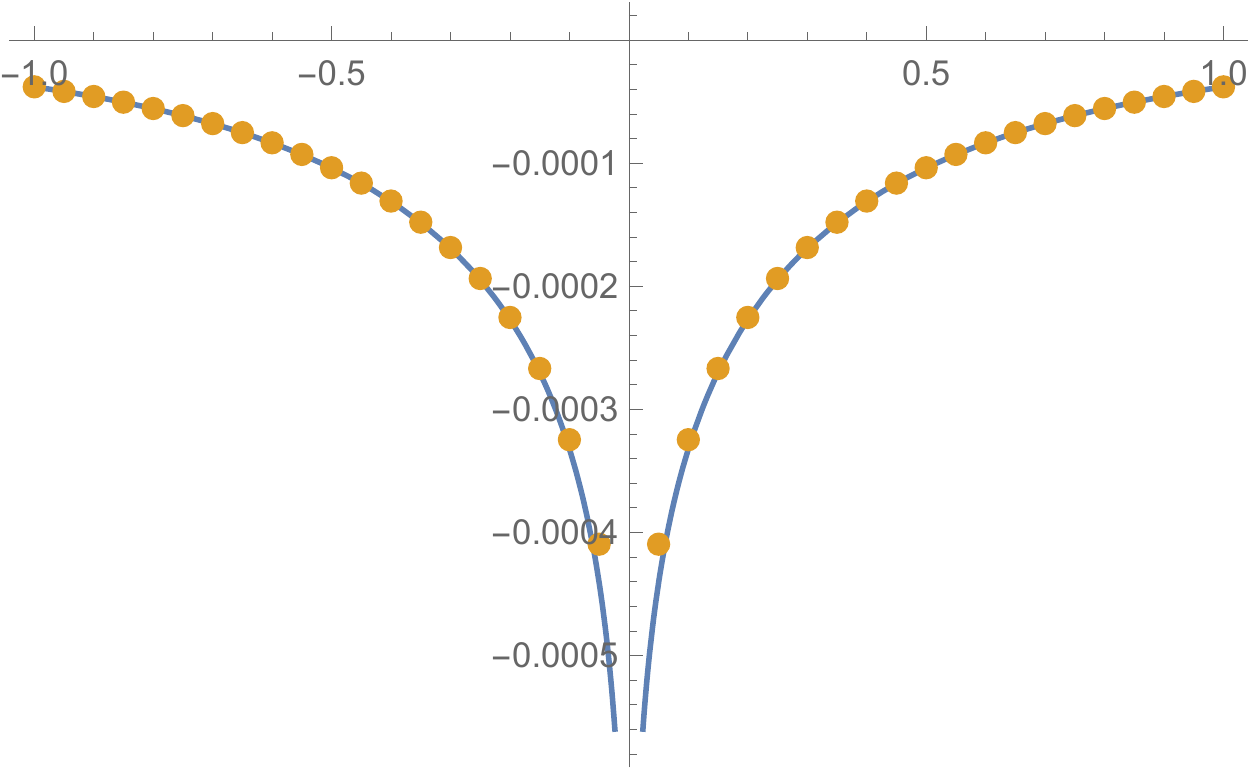}
\caption{$a = 0.999$, $v = 0$}
\end{subfigure}\\
\begin{subfigure}[t]{0.32\textwidth}
\centering
\includegraphics[width=\textwidth]{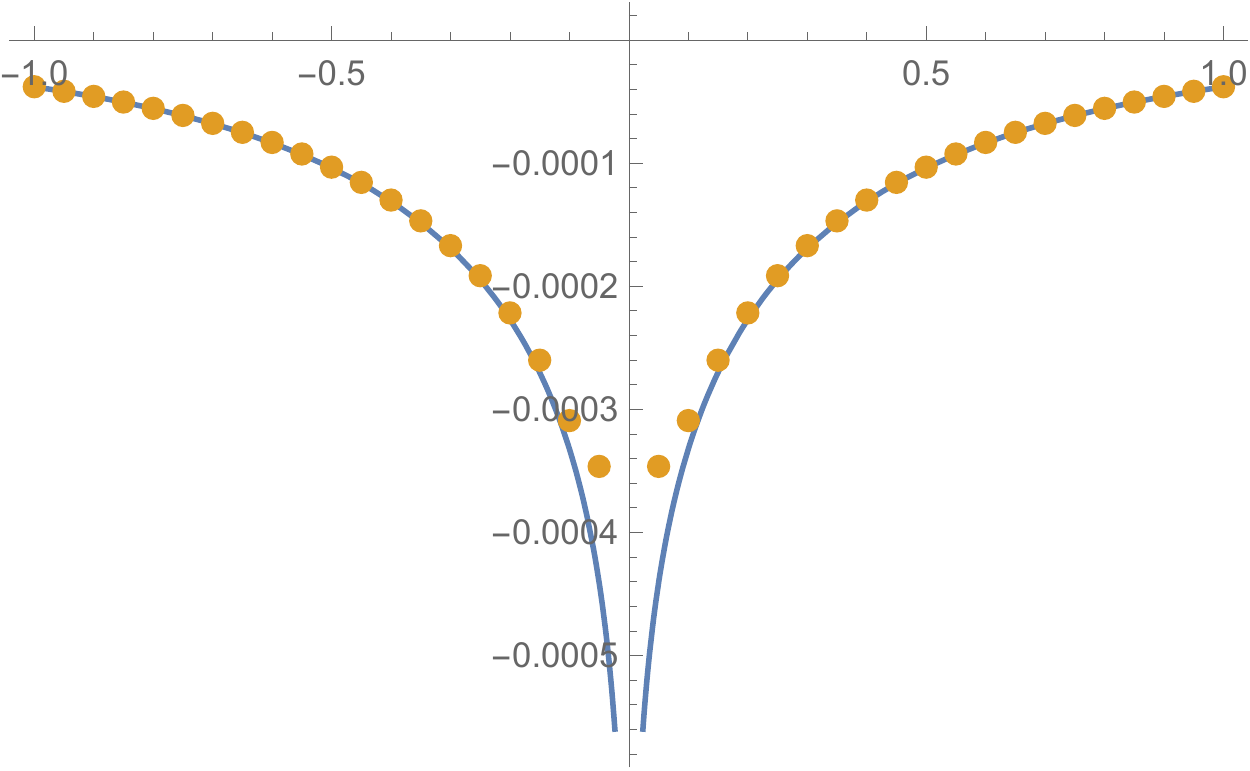}
\caption{$a = 0.999$, $v = 1$}
\end{subfigure}
\begin{subfigure}[t]{0.32\textwidth}
\centering
\includegraphics[width=\textwidth]{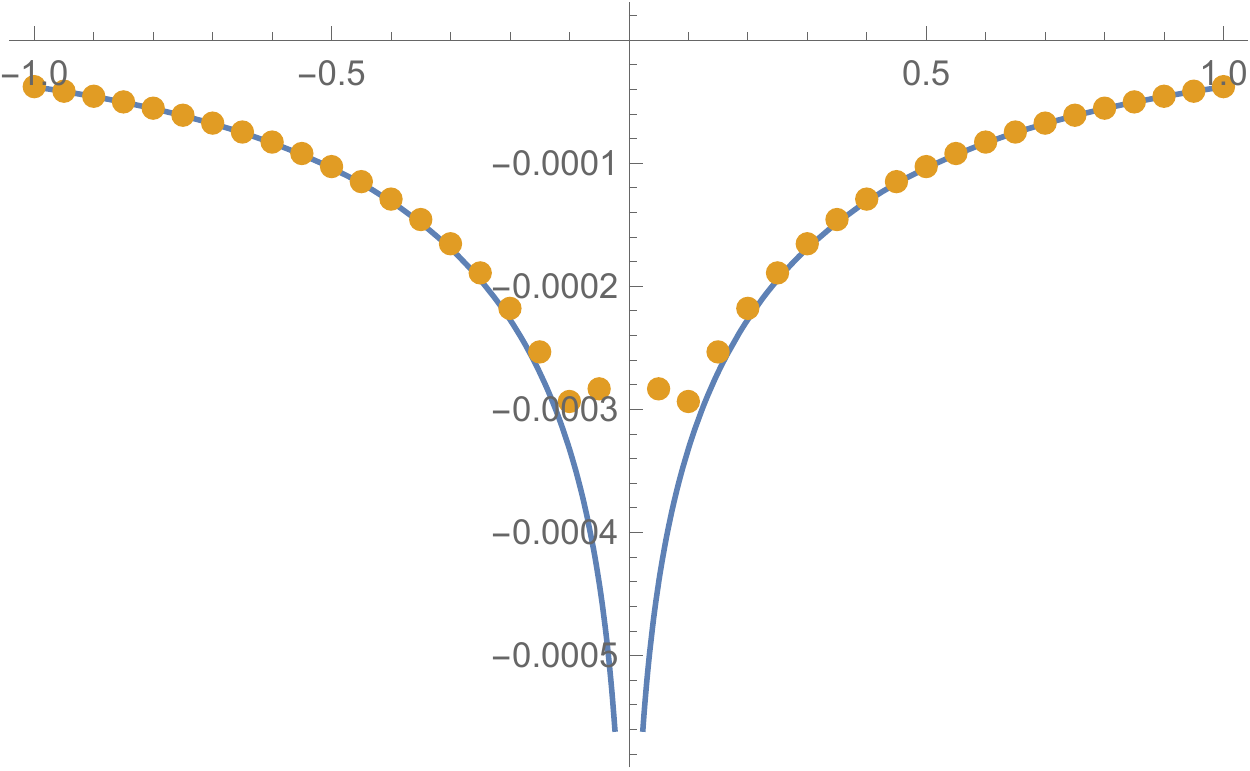}
\caption{$a = 0.999$, $v = 2$}
\end{subfigure}
\\
\begin{subfigure}[t]{0.32\textwidth}
\centering
\includegraphics[width=\textwidth]{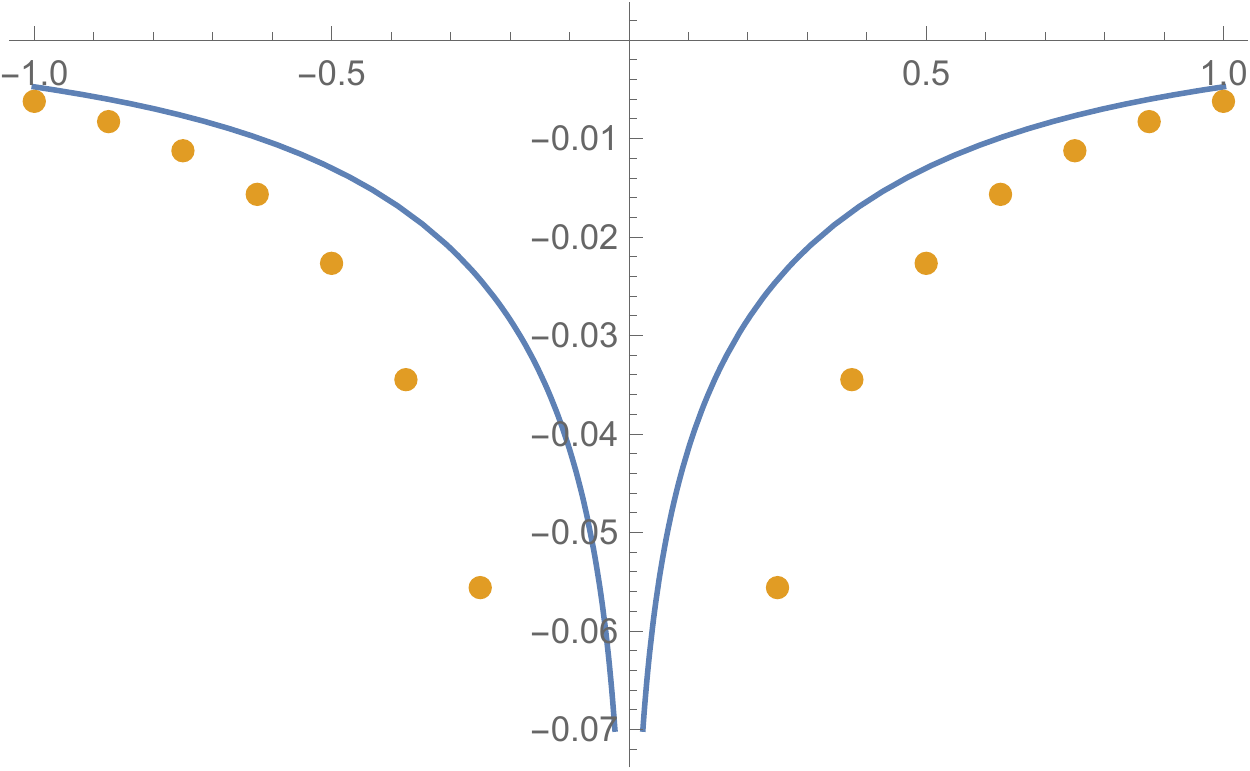}
\caption{$v = -2$}
\end{subfigure}
\begin{subfigure}[t]{0.32\textwidth}
\centering
\includegraphics[width=\textwidth]{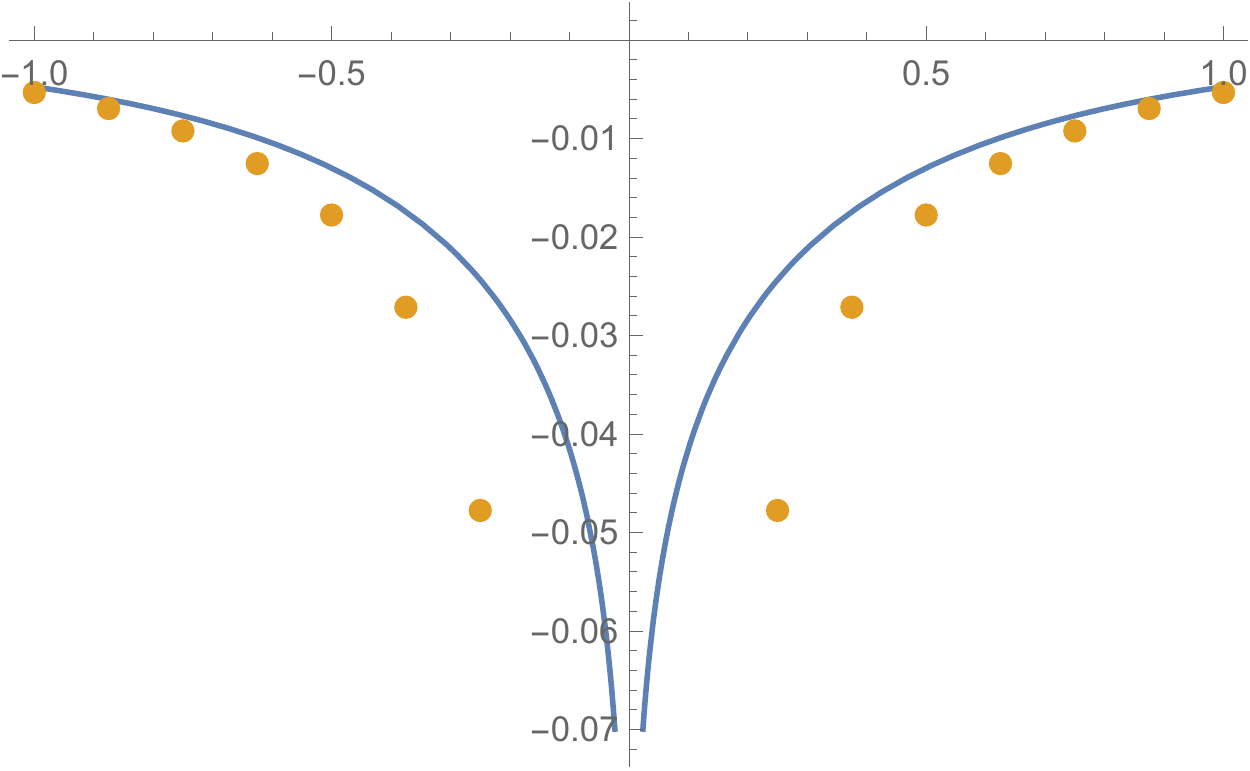}
\caption{$v = -1$}\label{fig:k11vm0}
\end{subfigure}
\begin{subfigure}[t]{0.32\textwidth}
\centering
\includegraphics[width=\textwidth]{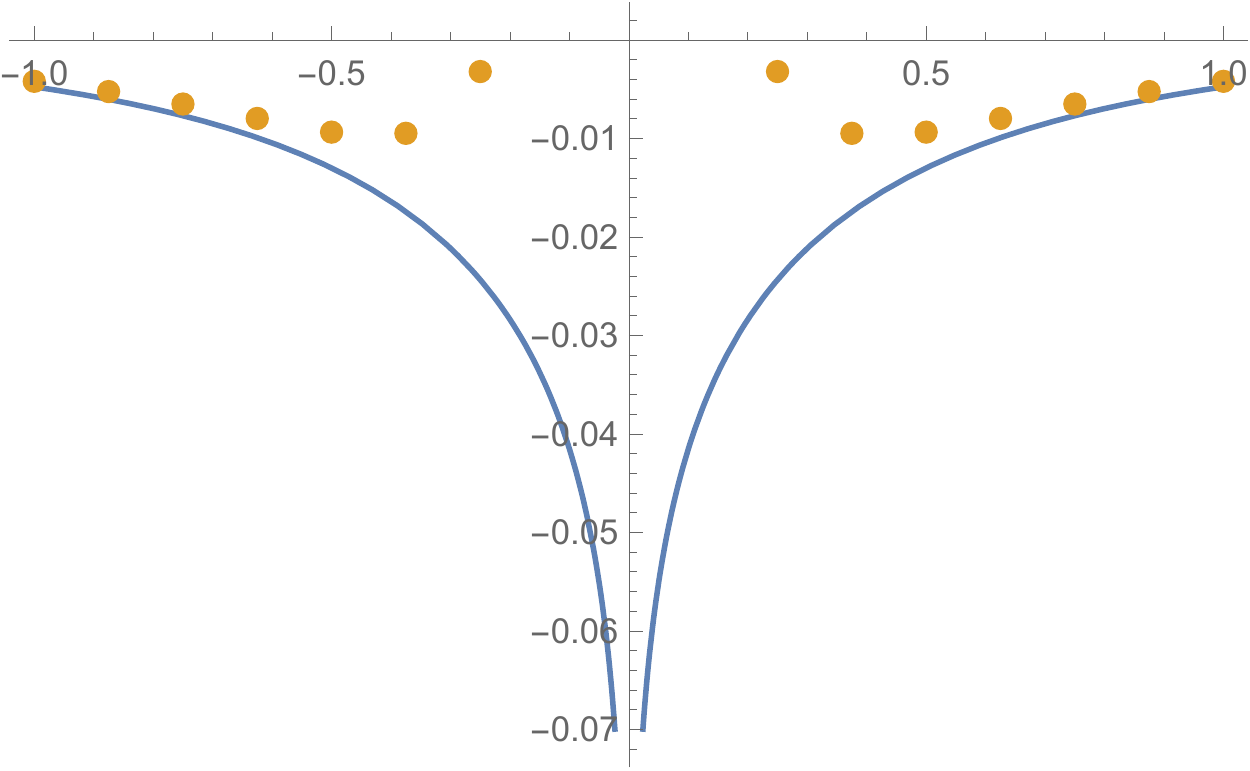}
\caption{$v = 0$}\label{fig:k11v0}
\end{subfigure}\\
\begin{subfigure}[t]{0.32\textwidth}
\centering
\includegraphics[width=\textwidth]{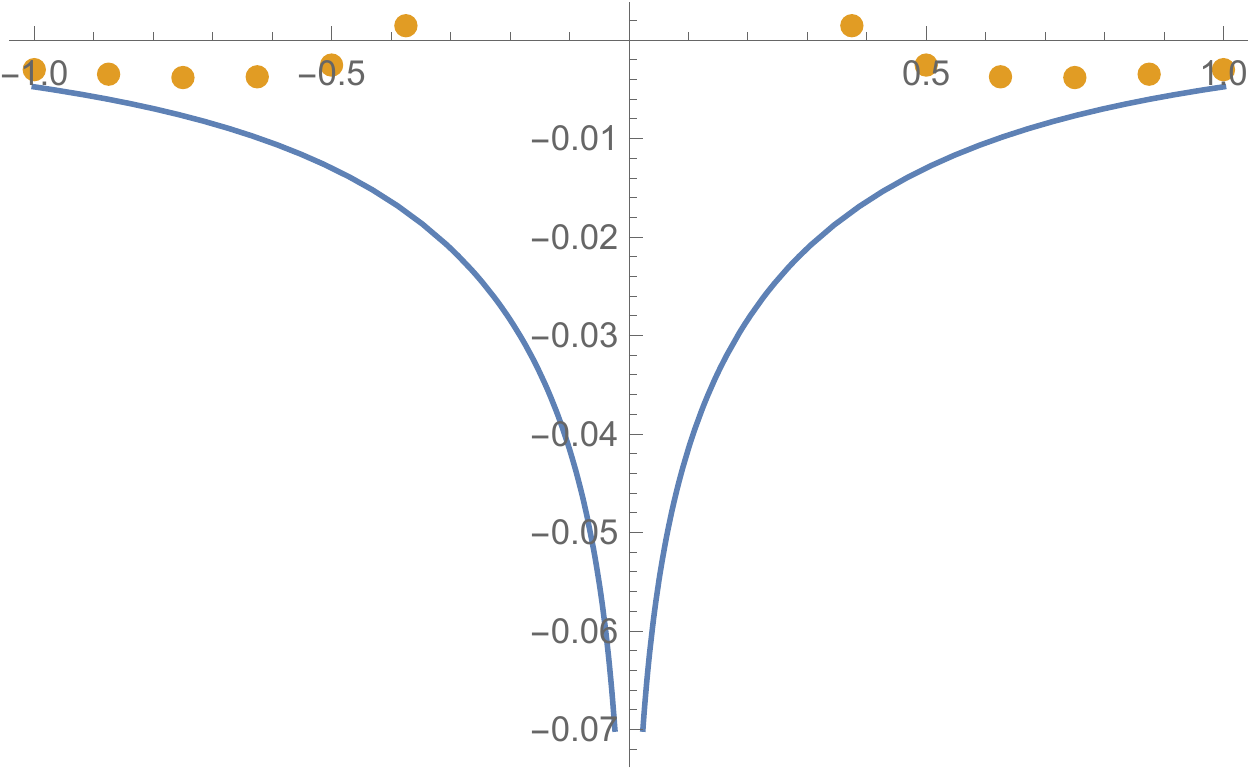}
\caption{$v = 1$}
\end{subfigure}
\begin{subfigure}[t]{0.32\textwidth}
\centering
\includegraphics[width=\textwidth]{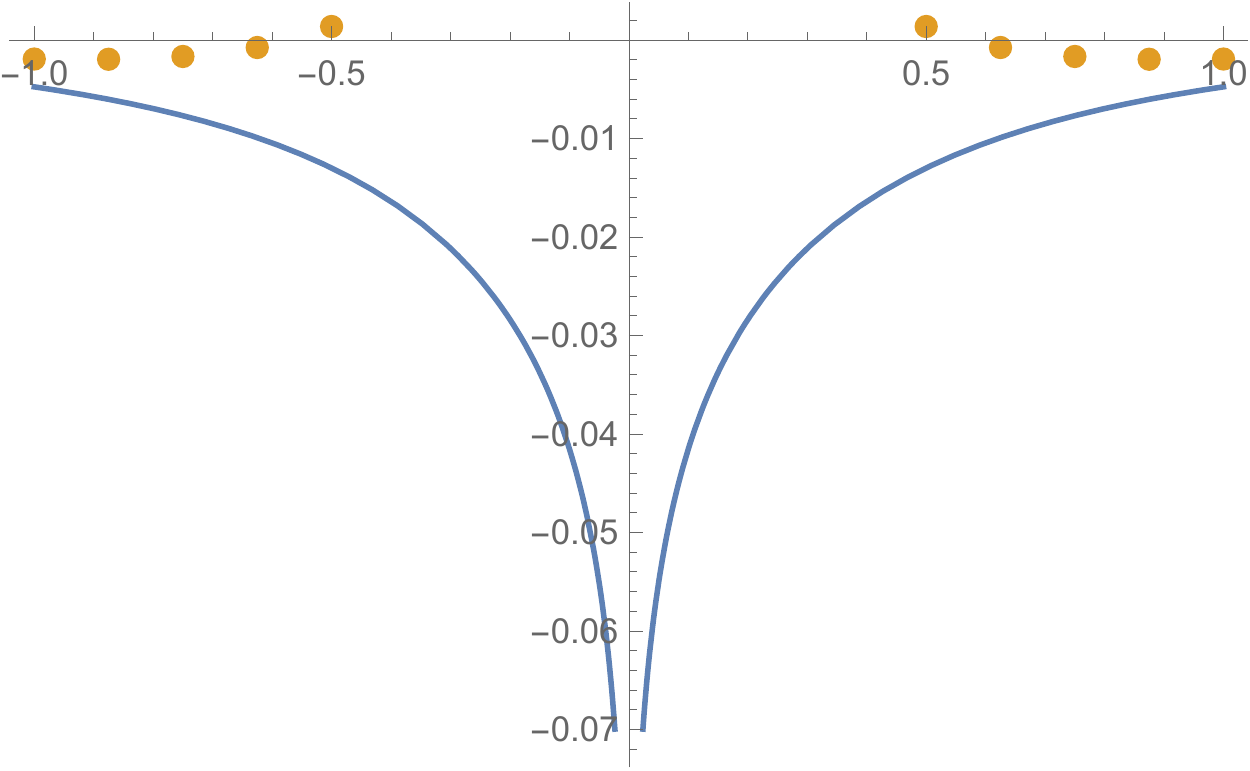}
\caption{$v = 2$}
\end{subfigure}
\caption{The blue line shows $(1-a) (-1)^{u+v} (-K_0(\sqrt{2}\bconst^2\alpha))/2)/\pi$ and the orange points show $i(-1)^{u+v}\mathbb{K}_{a,0,0}^{-1}(\mathbf{w}_0, \mathbf{b}_0 + 2ue_1 + 2ve_2)$, both plotted against $\alpha$, where $\alpha = (1-a) u$, for $a = 0.875$ and various values of $v$.}
\label{fig:k11numeric}
\end{figure}

To get an idea of the magnitude of the $o(m^{-1/2})$ terms in Equation~\ref{eq:Kasym}, we numerically evaluate the integral given in Equation~\ref{eq:translationinvariantoriginal}, and compare it to the leading order term given in Equation~\ref{eq:Kasym} for different values of $a$. In Figure~\ref{fig:k01numeric} we plot $(-1)^{u+v}\mathbb{K}_{a,0,0}^{-1}(\mathbf{w}_0, \mathbf{b}_1 + 2ue_1 + 2ve_2)$ against $h (-1)^{u+v} (-K_0(\sqrt{2}\bconst^2\alpha)/2 + \operatorname{sgn}(\alpha)K_1(\sqrt{2}\bconst^2\alpha)/\sqrt{2})/\pi$, and in Figure~\ref{fig:k11numeric} we plot $(-1)^{u+v}\mathbb{K}_{a,0,0}^{-1}(\mathbf{w}_0, \mathbf{b}_0 + 2ue_1 + 2ve_2)$ against $ih (-1)^{u+v} (-K_0(\sqrt{2}\bconst^2\alpha))/2)/\pi$, where $h = 1-a$ and $\alpha = h u$, for $a = 0.999$ and $a = 0.875$, with $v=-2,-1,0,1$ and $2$. The latter value of $a=0.875$ is what we use for the simulations in Section~\ref{sec:simulations}. The range of $\alpha$ shown here is $[-1,1]$.

We see that for $a = 0.999$, the leading order asymptotic term agrees closely with the exact integral, but for $a = 0.875$ there are some fairly large discrepancies in some of the plots. Unfortunately we are not able to run simulations large enough that this $o(m^{-1/2})$ term is negligible. As a result, in the next section we will present two-point correlations corresponding to pairs of dominos where the discrepancy between the exact value of $\mathbb{K}_{a,0,0}^{-1}(x, y)$ and the leading order asymptotic term are not too big, as in Figures~\ref{fig:k01v0}, \ref{fig:k11vm0} and \ref{fig:k11v0}.

\section{Simulation results}\label{sec:simulations}

\begin{figure}[htb]
\centering
\includegraphics[width=0.5\textwidth]{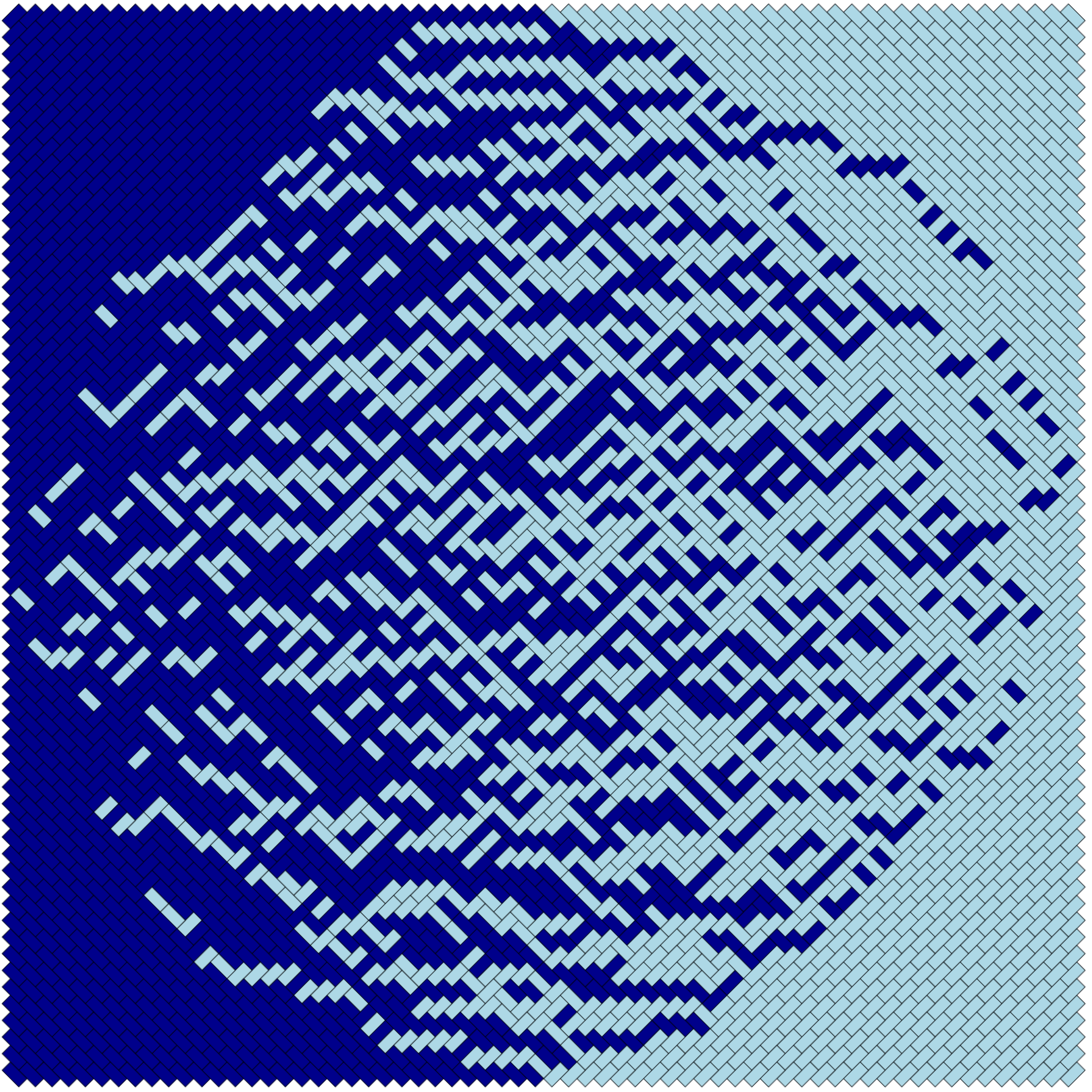}
\caption{Tiling of two-periodic weighted Aztec diamond with $n=64$ and $a = 0.8$}
\label{fig:dominotiling}
\end{figure}

We used Markov chain sampling to produce a large number of sample tilings (which are in bijection with dimer configurations) from the correct probability distribution. One such tiling is shown in Figure~\ref{fig:dominotiling}. We used the source code developed by Keating and Sridhar \cite{keating2018code} described in \cite{keating2018}, with some modifications. We ran our code on a GTX 1080 Ti GPU.

Here we will show the experimental two-point correlations for pairs of dimers along the diagonal with one fixed and one variable, for some different types of dimers. 

To simplify notation, we make the following definition. Take $x \in \mathtt{W}_\e1$ and $y \in \mathtt{B}_\e2$ with $\e1,\e2\in \{0,1\}$. For $-1/\sqrt{2} \leq \alpha_x,\, \alpha_y < 0$ define
\begin{multline}
q_{\e1 \e2}(\alpha_x,\alpha_y) = \bigg(-\frac{1}{2\pi}K_0(\sqrt{2}\bconst^2|\alpha_y - \alpha_x|) \\
+ (\e2-\e1) \operatorname{sgn}(\alpha_y-\alpha_x)\frac{1}{\sqrt{2}\pi}K_1(\sqrt{2}\bconst^2|\alpha_y - \alpha_x|)\bigg) + \psi(\alpha_x, \alpha_y,\e1,\e2)
\end{multline} and for $\alpha_x < -1/\sqrt{2}$ or $\alpha_x < -1/\sqrt{2}$ define
\begin{multline}
q_{\e1 \e2}(\alpha_x,\alpha_y) = \bigg(-\frac{1}{2\pi}K_0(\sqrt{2}\bconst^2|\alpha_y - \alpha_x|) \\+ (\e2-\e1) \operatorname{sgn}(\alpha_y-\alpha_x)\frac{1}{\sqrt{2}\pi}K_1(\sqrt{2}\bconst^2|\alpha_y - \alpha_x|)\bigg) + \frac{I_0(\alpha_x, \alpha_y,  \e1, \e2)}{4\pi}
+ \psi(\alpha_x, \alpha_y, \e1,\e2).
\end{multline} Then Conjecture~\ref{conj:mainresult} can be written
\begin{equation}\label{eq:Kqconj}
K_a^{-1}(x,y) =  \bconst m^{-1/2} \frac{\zeta(x,y)}{\Sigma(x,y)} q_{\e1 \e2}(\alpha_x,\alpha_y) + o(m^{-1/2}).\end{equation}

Now consider two edges $e = (x,y)$ and $\widetilde{e} = (\widetilde{x},\widetilde{y})$ with $x,\widetilde{x} \in \mathtt{W}$ and $y, \widetilde{y}\in\mathtt{B}$. From Equation~\ref{eq:rho_twopoint}, the two-point correlation $\rho(e, \widetilde{e})$ between these two edges is given by
\[
\rho(e, \widetilde{e}) = K_a(y,x) K_a(\widetilde{y},\widetilde{x}) (K_a^{-1}(x,y)K_a^{-1}(\widetilde{x},\widetilde{y}) - K_a^{-1}(x,\widetilde{y})K_a^{-1}(\widetilde{x},y)),
\] and the covariance between the two edges is given by \[
\cov(e,\widetilde{e}) = \rho(e, \widetilde{e}) - \rho(e)\rho(\widetilde{e}) = - K_a(y,x) K_a(\widetilde{y},\widetilde{x}) K_a^{-1}(x,\widetilde{y})K_a^{-1}(\widetilde{x},y)
\] where $\rho(e)$ is the one-point correlation function of $e$, i.e. the probability that a randomly chosen dimer configuration contains edge $e$. We compare the experimental covariance between edges to our conjectured asymptotics. 

Suppose $e = (x,y)$ and $\widetilde{e} = (\widetilde{x},\widetilde{y})$ with $x \in \mathtt{W}_\e1$, $\widetilde{x} \in \mathtt{W}_{\widetilde{\e1}}$ and $y \in \mathtt{B}_\e2$, $\widetilde{y}\in\mathtt{B}_{\widetilde{\e2}}$. Suppose further that these dimers lie near the diagonal as in Equation~\ref{eq:asymptoticcoord}. Let $\alpha$ be the asymptotic coordinate of $x$ and $y$, and let $\widetilde{\alpha}$ be the asymptotic coordinate of $\widetilde{x}$ and $\widetilde{y}$. We recall that $K_a(y,x) = \Sigma(x,y) + O(h)$ where $h = 1-a$. Then from Equation~\ref{eq:Kqconj} we can conjecture \[
\cov(e,\widetilde{e}) = - \bconst^2 m^{-1} \frac{\Sigma(x, y) \Sigma(\widetilde{x},\widetilde{y})}{\Sigma(x,\widetilde{y})\Sigma(\widetilde{x},y)}  \zeta(x,\widetilde{y}) \zeta(\widetilde{x},y) q_{\e1 \widetilde{\e2}}(\alpha,\widetilde{\alpha}) q_{\widetilde{\e1} \e2}(\widetilde{\alpha},\alpha) + o(m^{-1}).
\]

\begin{figure}[ht]
\centering
\begin{subfigure}[t]{0.47\textwidth}
\centering
\includegraphics[width=\textwidth]{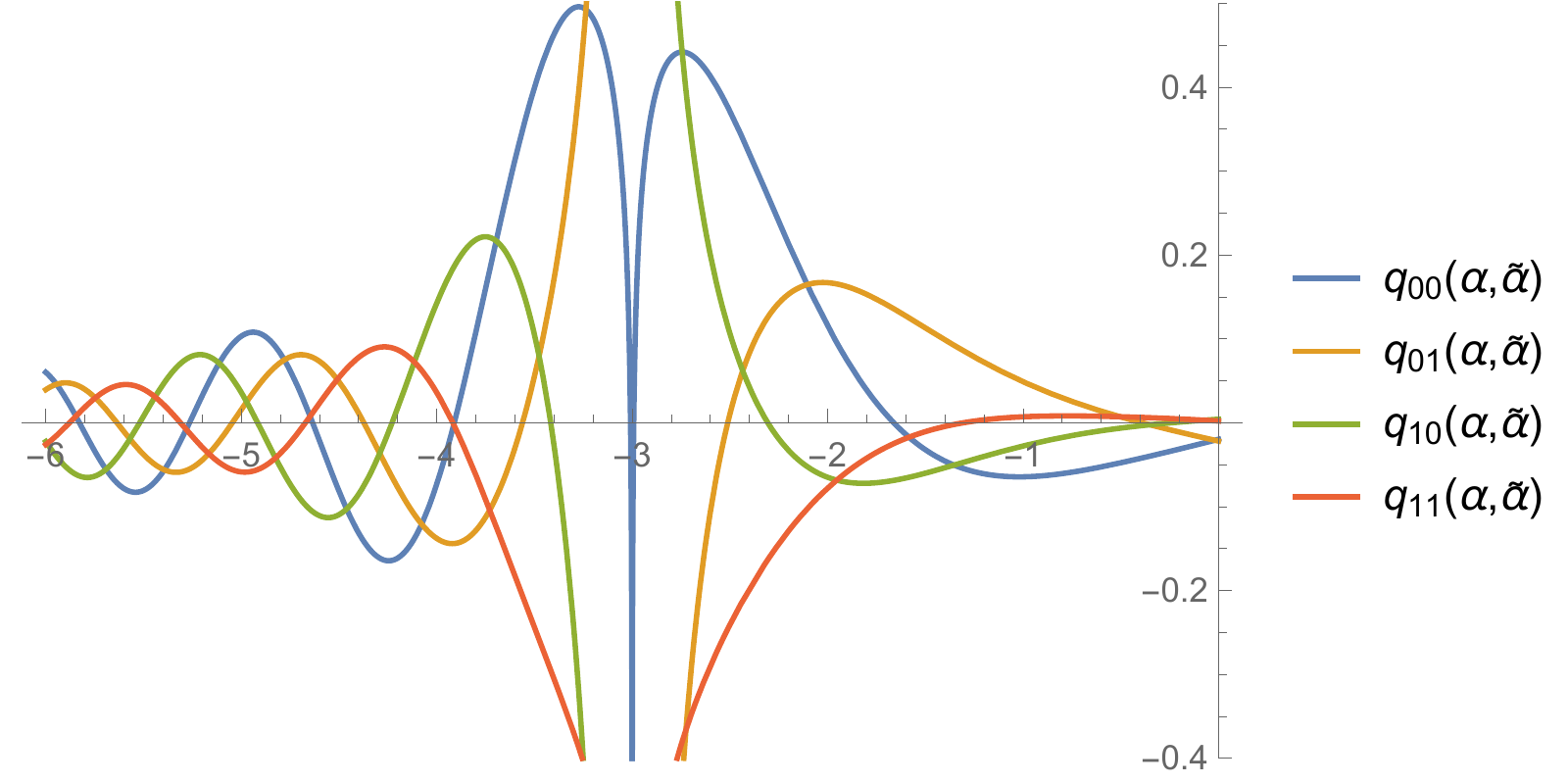}
\caption{$\widetilde{\alpha} = -3$}
\end{subfigure}\hfil
\begin{subfigure}[t]{0.47\textwidth}
\centering
\includegraphics[width=\textwidth]{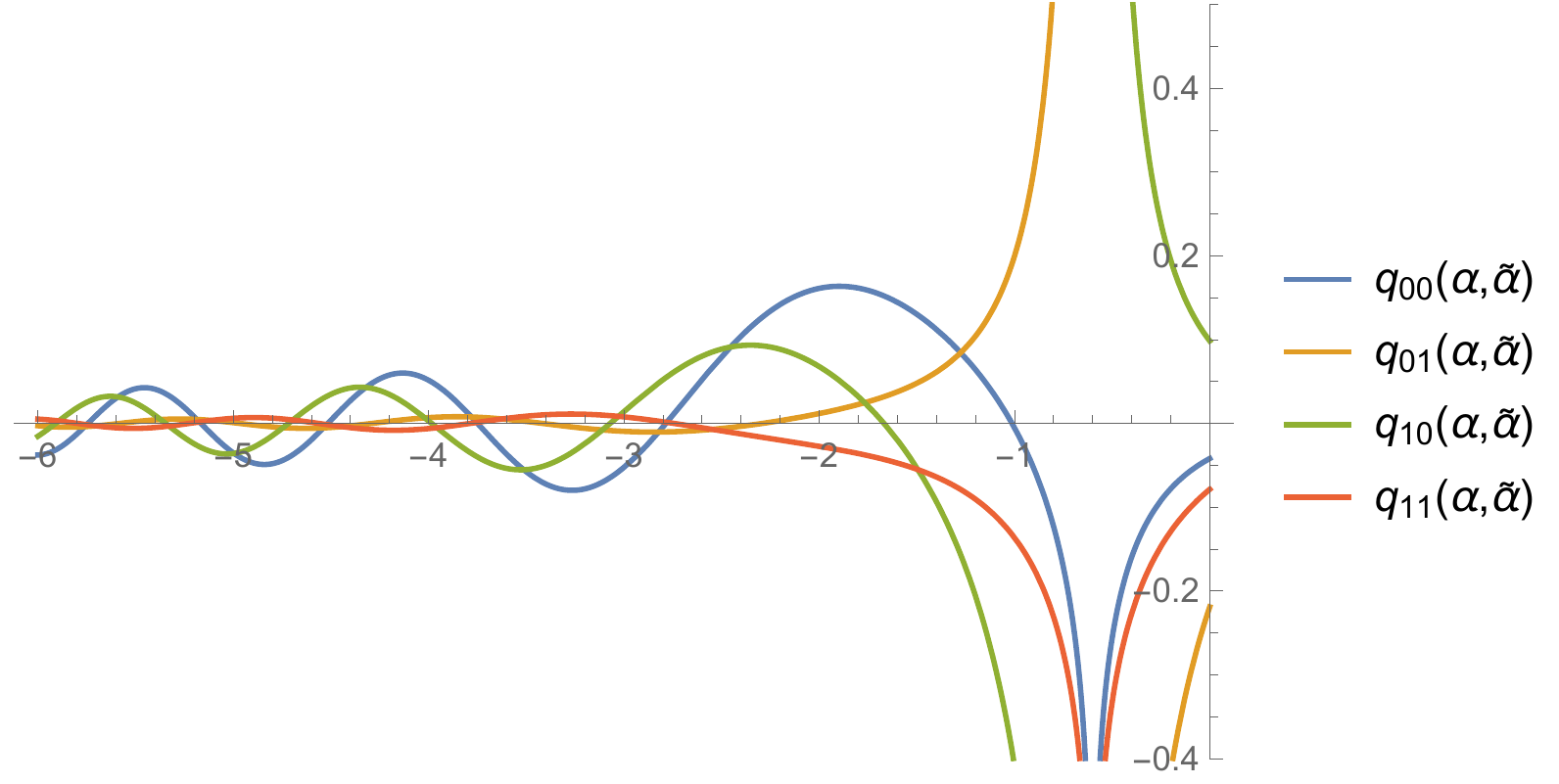}
\caption{$\widetilde{\alpha} = -0.6$}
\end{subfigure}
\caption{Plots of $q_{\e1\e2}(\alpha,\widetilde{\alpha})$ against $\alpha$ for fixed $\widetilde{\alpha}$ and $\e1,\e2\in \{0,1\}$.}\label{fig:qs}
\end{figure}

We can show that $q_{0 0}(\alpha,\widetilde{\alpha})= q_{0 0}(\widetilde{\alpha},\alpha)$, $q_{1 1}(\alpha,\widetilde{\alpha})= q_{11}(\widetilde{\alpha},\alpha)$, $q_{0 1}(\alpha,\widetilde{\alpha})= q_{1 0}(\widetilde{\alpha},\alpha)$ and $q_{1 0}(\alpha,\widetilde{\alpha})= q_{0 1}(\widetilde{\alpha},\alpha)$. In Figure~\ref{fig:qs} we plot these quantities for $\widetilde{\alpha} = -3, -0.6$.

\begin{figure}
\centering
\begin{subfigure}[t]{0.32\textwidth}
\centering
\includegraphics[width=\textwidth]{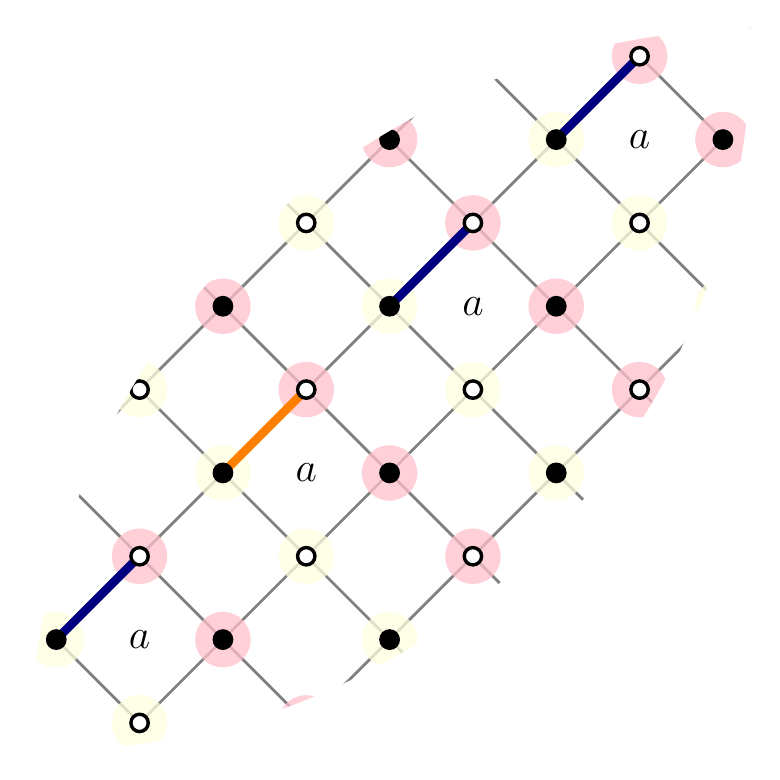}
\caption{$\e1 = 1, \e2 = 0, \widetilde{\e1}=1, \widetilde{\e2}=0$}
\end{subfigure}
\begin{subfigure}[t]{0.32\textwidth}
\centering
\includegraphics[width=\textwidth]{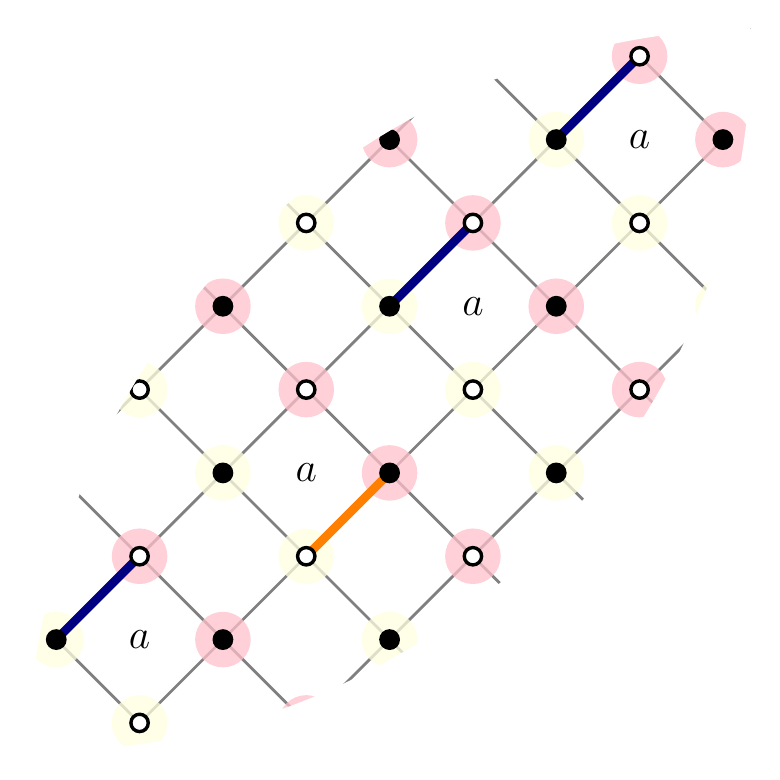}
\caption{$\e1 = 1, \e2 = 0, \widetilde{\e1}=0, \widetilde{\e2}= 1$}
\end{subfigure}
\begin{subfigure}[t]{0.32\textwidth}
\centering
\includegraphics[width=\textwidth]{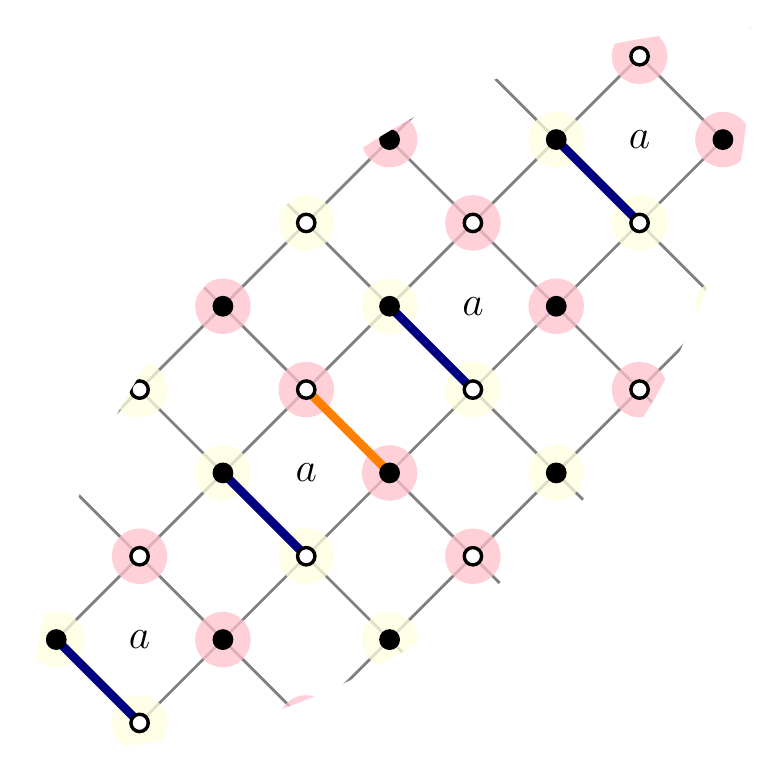}
\caption{$\e1 = 0, \e2 = 0, \widetilde{\e1}=1, \widetilde{\e2}=1$}
\end{subfigure}\\
\begin{subfigure}[t]{0.32\textwidth}
\centering
\includegraphics[width=\textwidth]{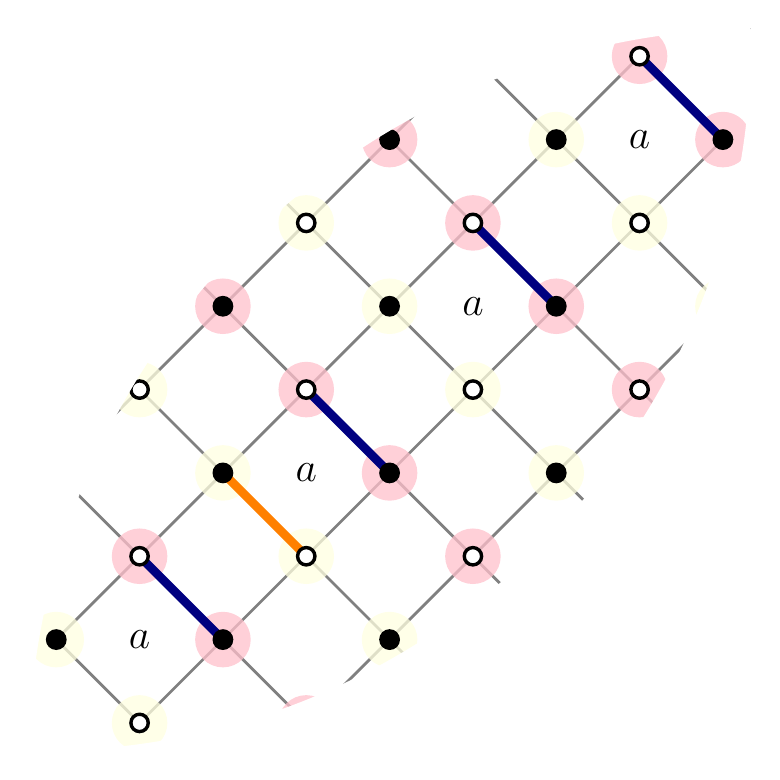}
\caption{$\e1 = 1, \e2 = 1, \widetilde{\e1}=0, \widetilde{\e2}=0$}
\end{subfigure}
\begin{subfigure}[t]{0.32\textwidth}
\centering
\includegraphics[width=\textwidth]{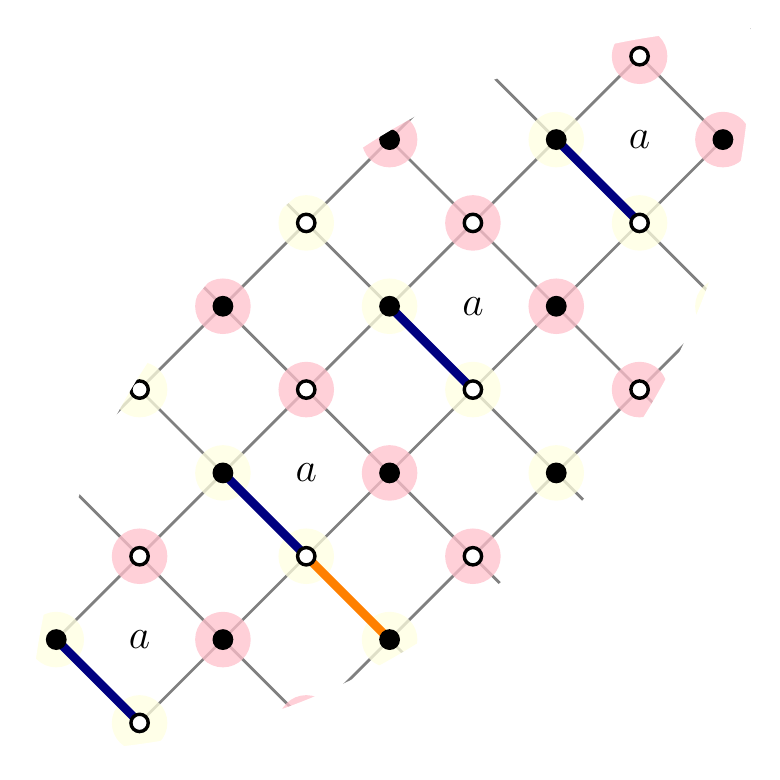}
\caption{$\e1 = 0, \e2 = 0, \widetilde{\e1}=0, \widetilde{\e2}=0$}
\end{subfigure}
\begin{subfigure}[t]{0.32\textwidth}
\centering
\includegraphics[width=\textwidth]{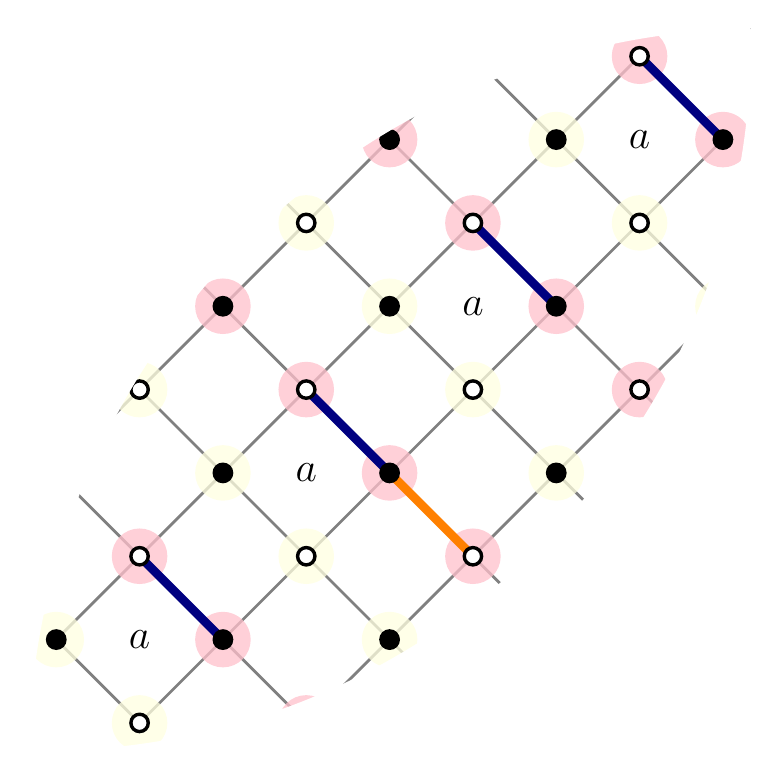}
\caption{$\e1 = 1, \e2 = 1, \widetilde{\e1}=1, \widetilde{\e2}=1$}
\end{subfigure}
\caption{Six different choices for $e$, shown in navy, and $\widetilde{e}$, shown in orange. Note that $\widetilde{e}$ is fixed while $e$ moves along the diagonal. Vertices in $\mathtt{W}_0\cup \mathtt{B}_0$ are colored in yellow, while vertices in $\mathtt{W}_1\cup\mathtt{B}_1$ are colored in pink.}\label{fig:dimerpairs}
\end{figure}

\begin{figure}
\centering
\begin{subfigure}[t]{0.45\textwidth}
\centering
\includegraphics[width=\textwidth]{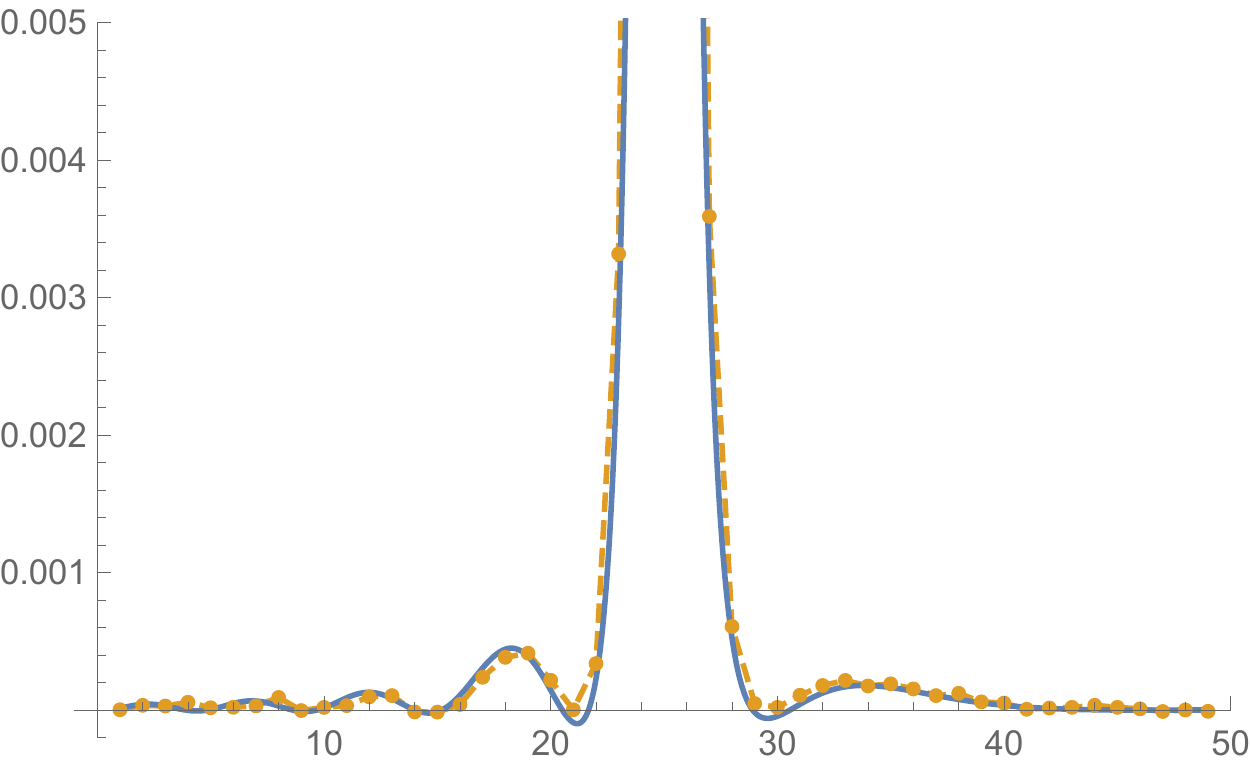}
\caption{$\e1 = 1, \e2 = 0, \widetilde{\e1}=1, \widetilde{\e2}=0$}
\end{subfigure}
\begin{subfigure}[t]{0.45\textwidth}
\centering
\includegraphics[width=\textwidth]{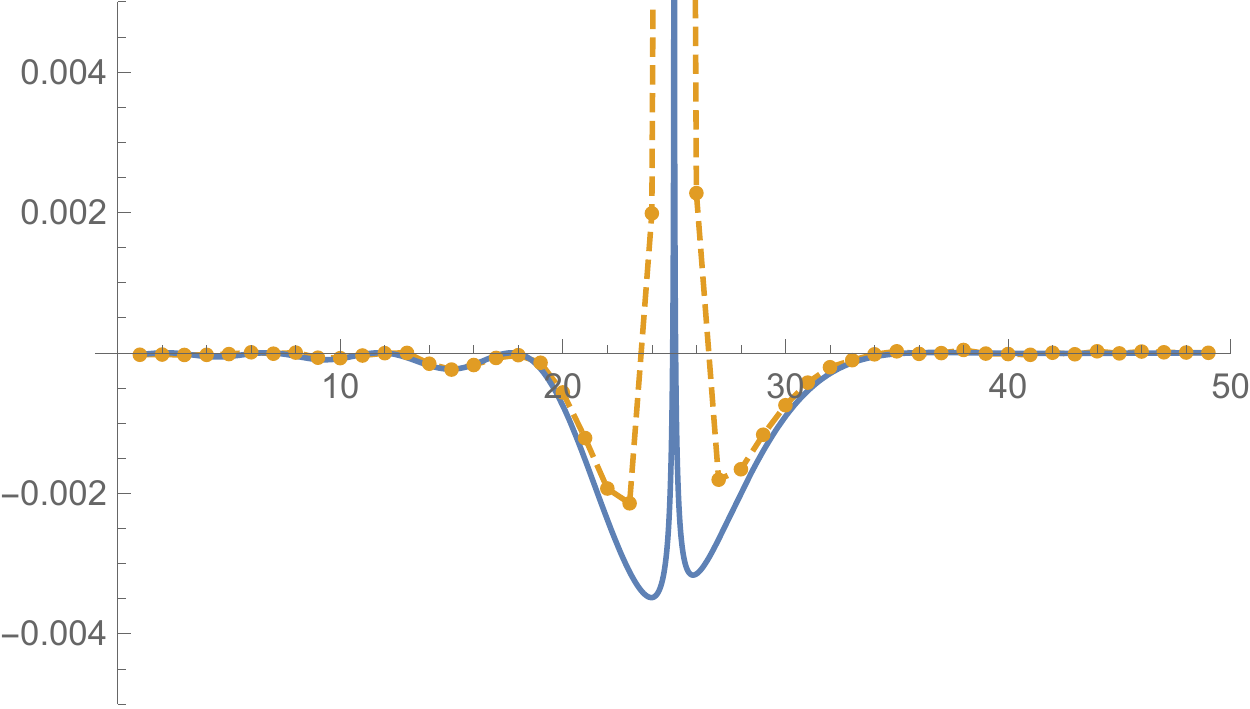}
\caption{$\e1 = 1, \e2 = 0, \widetilde{\e1}=0, \widetilde{\e2}= 1$}
\end{subfigure}\\
\begin{subfigure}[t]{0.45\textwidth}
\centering
\includegraphics[width=\textwidth]{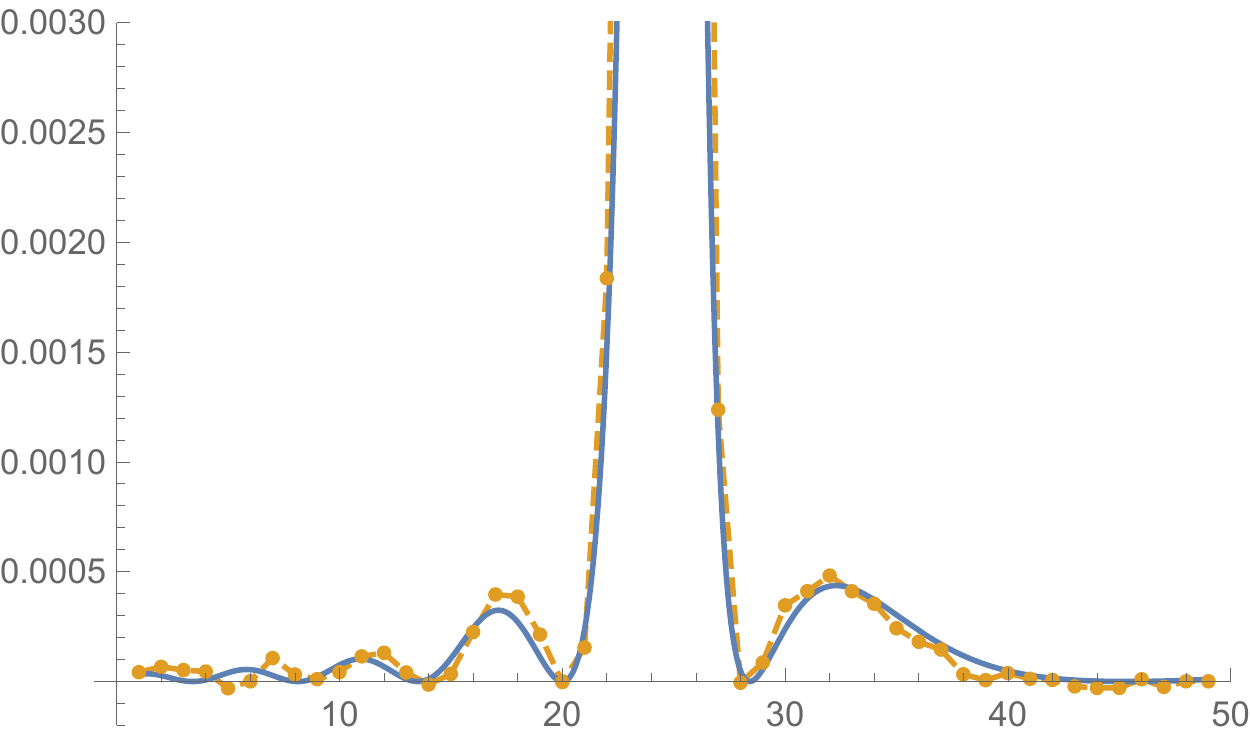}
\caption{$\e1 = 0, \e2 = 0, \widetilde{\e1}=1, \widetilde{\e2}=1$}
\end{subfigure}
\begin{subfigure}[t]{0.45\textwidth}
\centering
\includegraphics[width=\textwidth]{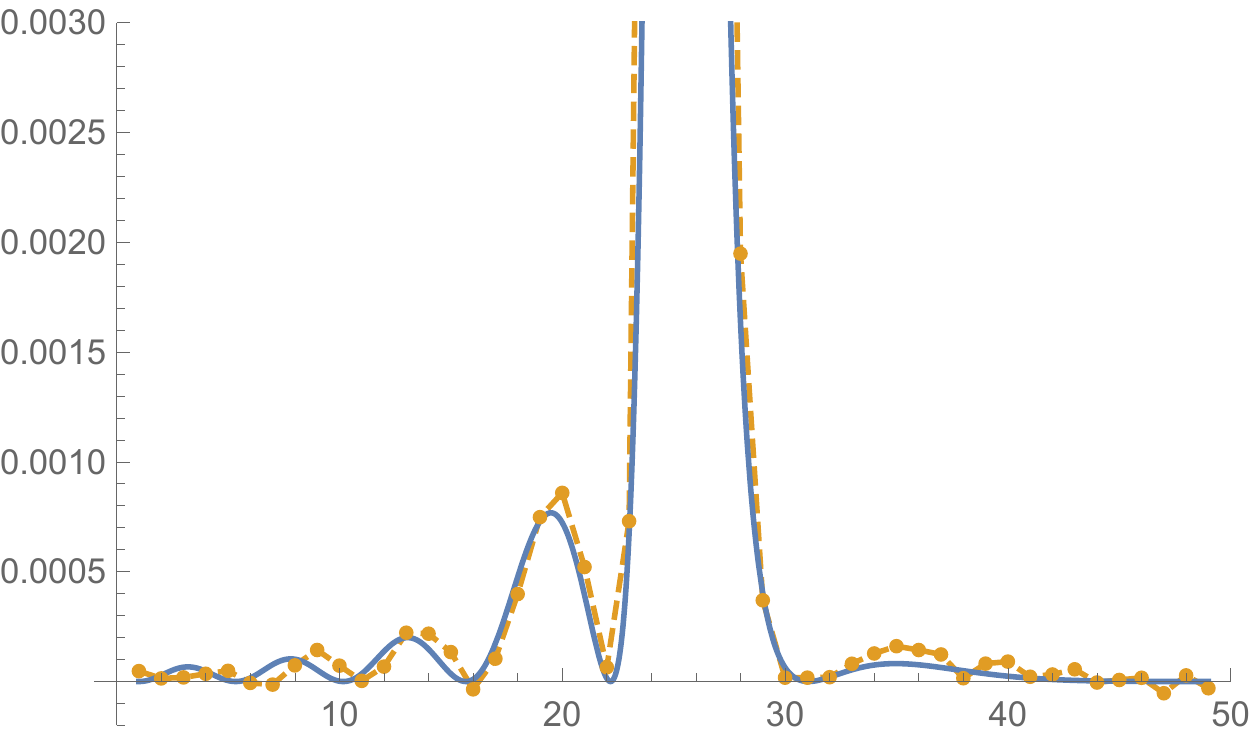}
\caption{$\e1 = 1, \e2 = 1, \widetilde{\e1}=0, \widetilde{\e2}=0$}
\end{subfigure}\\
\begin{subfigure}[t]{0.45\textwidth}
\centering
\includegraphics[width=\textwidth]{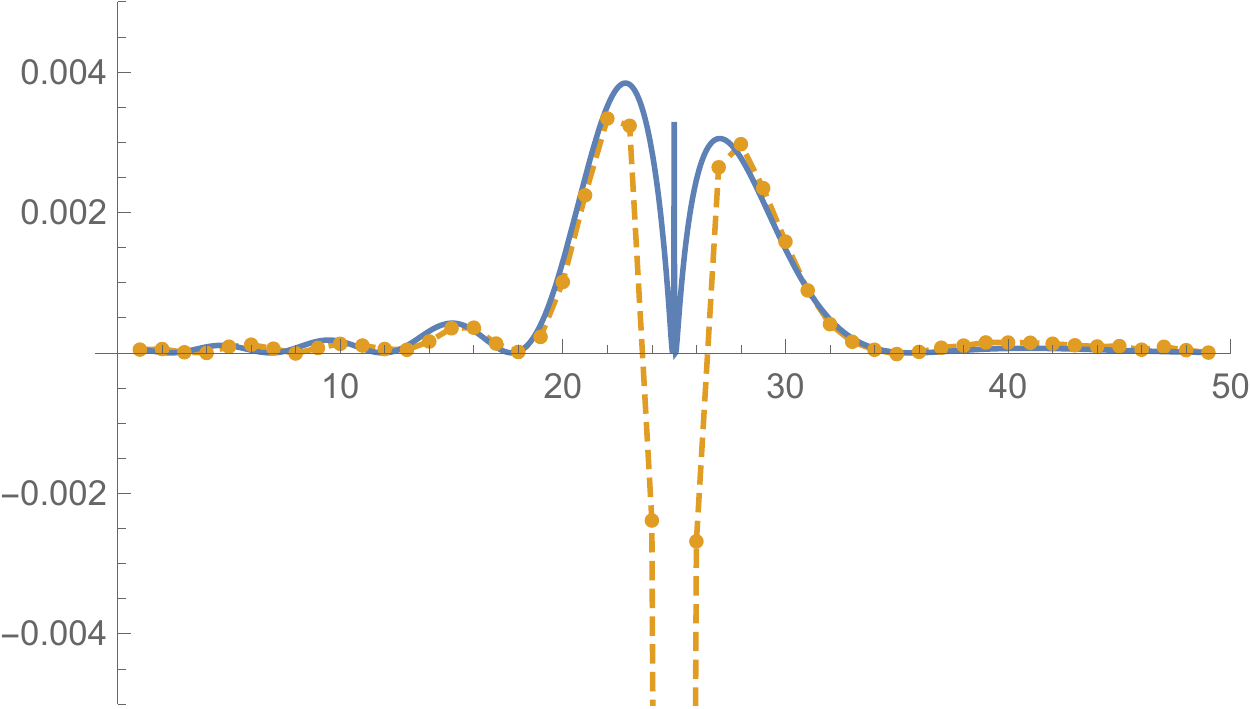}
\caption{$\e1 = 0, \e2 = 0, \widetilde{\e1}=0, \widetilde{\e2}=0$}
\end{subfigure}
\begin{subfigure}[t]{0.45\textwidth}
\centering
\includegraphics[width=\textwidth]{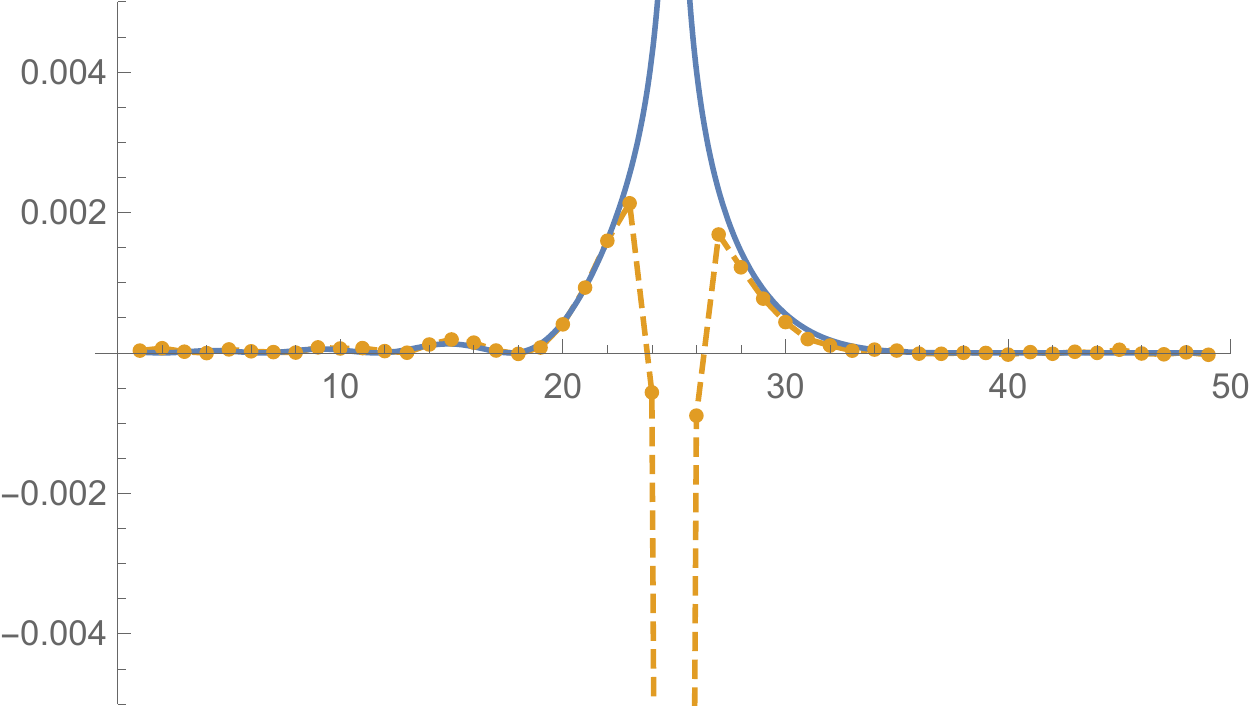}
\caption{$\e1 = 1, \e2 = 1, \widetilde{\e1}=1, \widetilde{\e2}=1$}
\end{subfigure}
\caption{$\widetilde{\alpha} = -3$. The blue solid lines show $-m^{-1}s(x,y,\widetilde{x},\widetilde{y}) q_{\e1 \widetilde{\e2}}(\alpha,\widetilde{\alpha}) q_{\widetilde{\e1} \e2}(\widetilde{\alpha},\alpha)$. The orange markers connected by dashed lines show the experimental covariances $\cov(e,\widetilde{e})$, for edges $e,\widetilde{e}$ as shown in Figure~\ref{fig:dimerpairs}. Here, $n=256$ and $a=0.875$.}\label{fig:expalpha3}
\end{figure}

\begin{figure}
\centering
\begin{subfigure}[t]{0.45\textwidth}
\centering
\includegraphics[width=\textwidth]{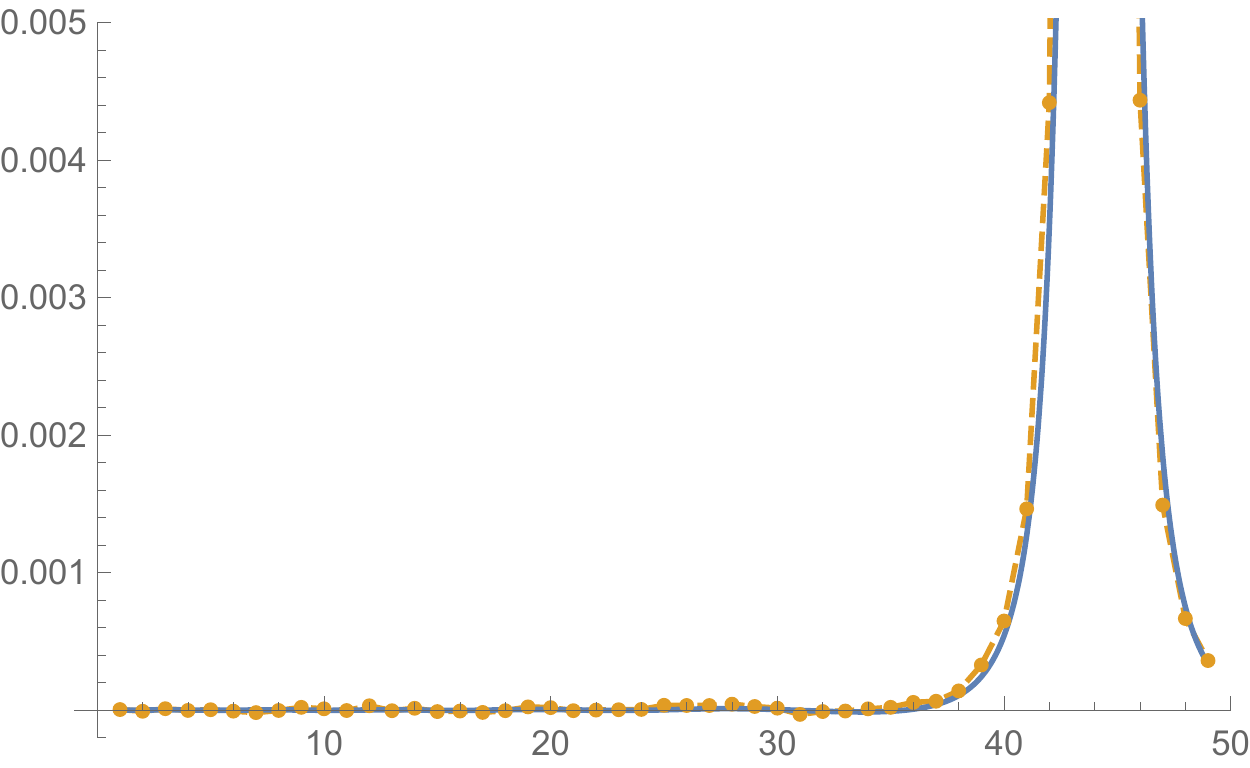}
\caption{$\e1 = 1, \e2 = 0, \widetilde{\e1}=1, \widetilde{\e2}=0$}
\end{subfigure}
\begin{subfigure}[t]{0.45\textwidth}
\centering
\includegraphics[width=\textwidth]{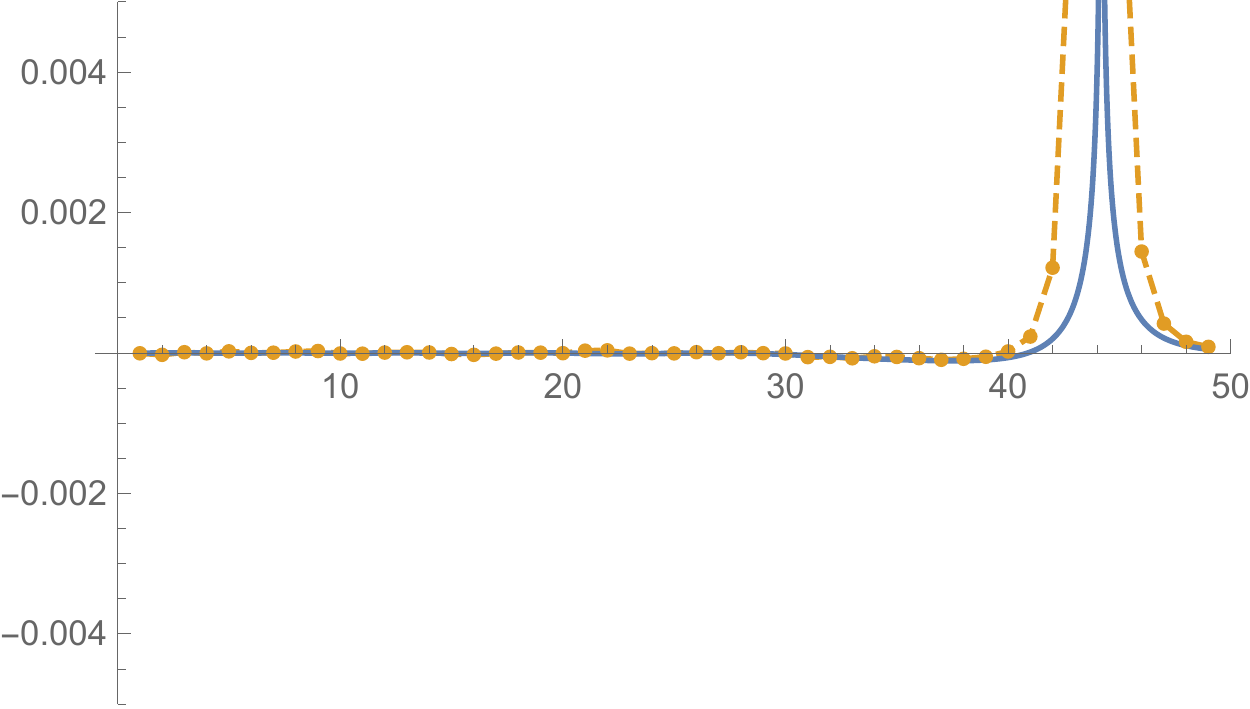}
\caption{$\e1 = 1, \e2 = 0, \widetilde{\e1}=0, \widetilde{\e2}= 1$}
\end{subfigure}\\
\begin{subfigure}[t]{0.45\textwidth}
\centering
\includegraphics[width=\textwidth]{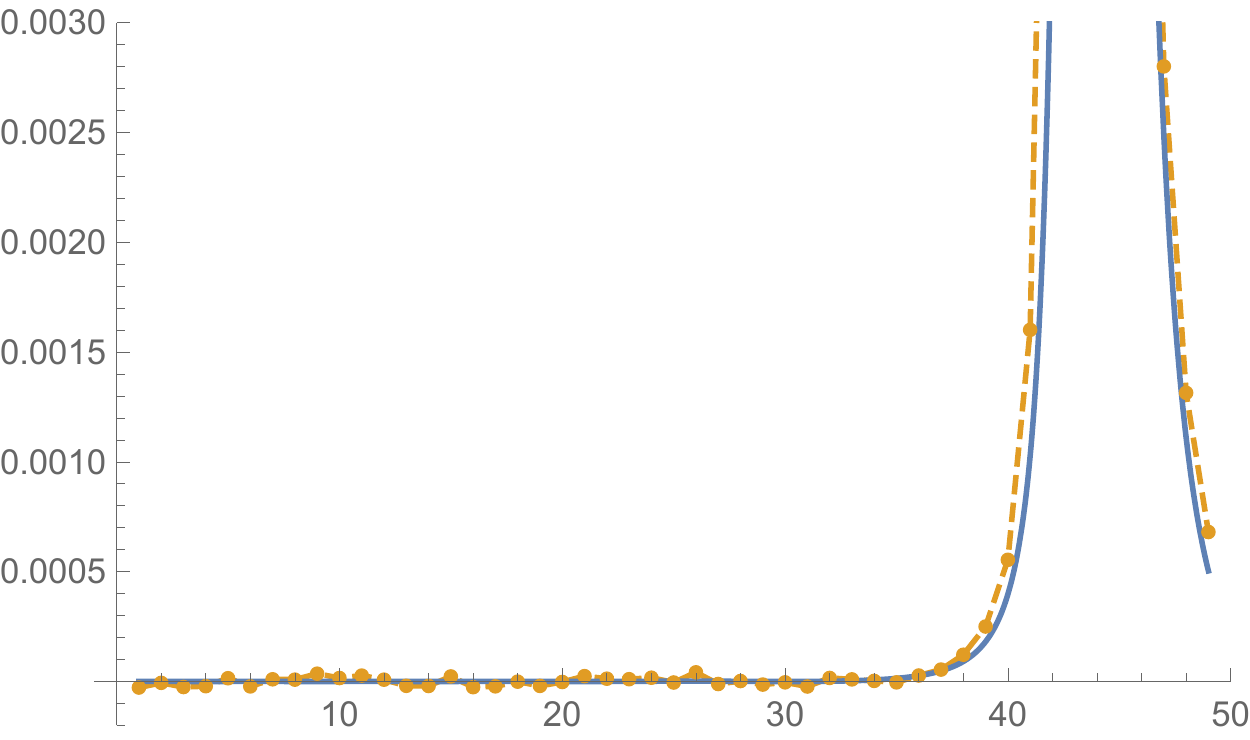}
\caption{$\e1 = 0, \e2 = 0, \widetilde{\e1}=1, \widetilde{\e2}=1$}
\end{subfigure}
\begin{subfigure}[t]{0.45\textwidth}
\centering
\includegraphics[width=\textwidth]{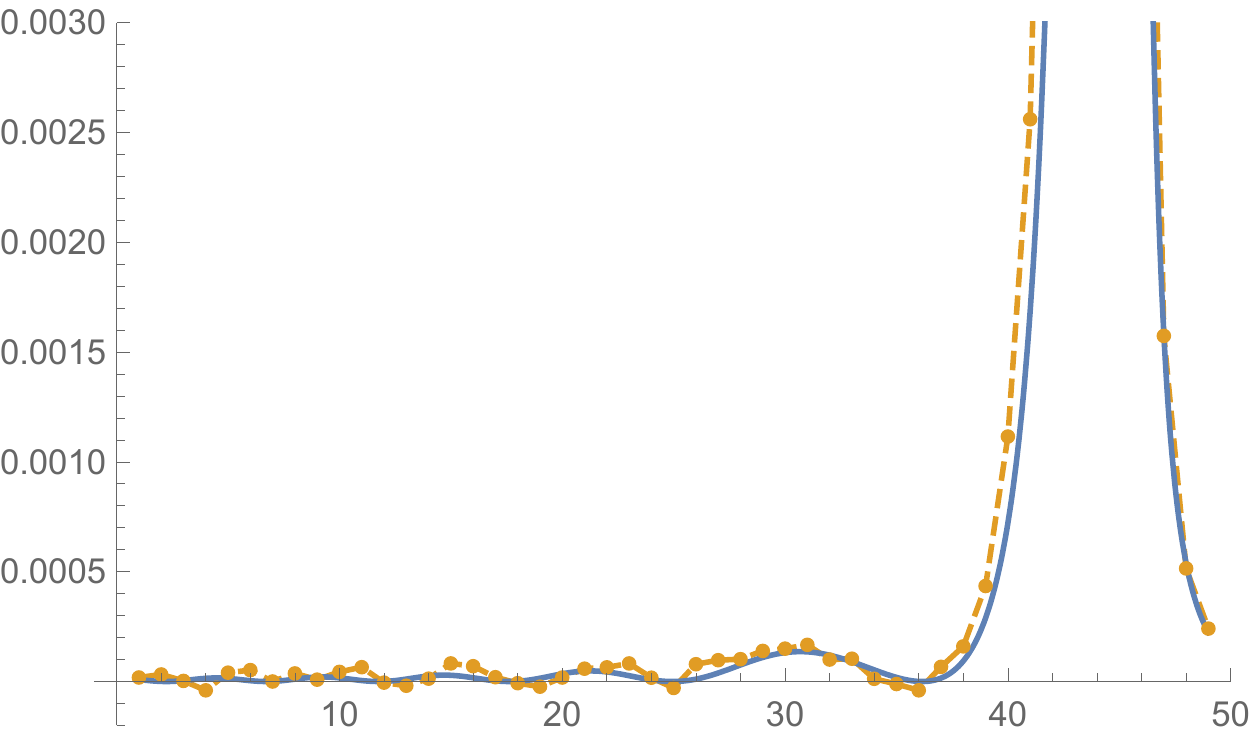}
\caption{$\e1 = 1, \e2 = 1, \widetilde{\e1}=0, \widetilde{\e2}=0$}
\end{subfigure}\\
\begin{subfigure}[t]{0.45\textwidth}
\centering
\includegraphics[width=\textwidth]{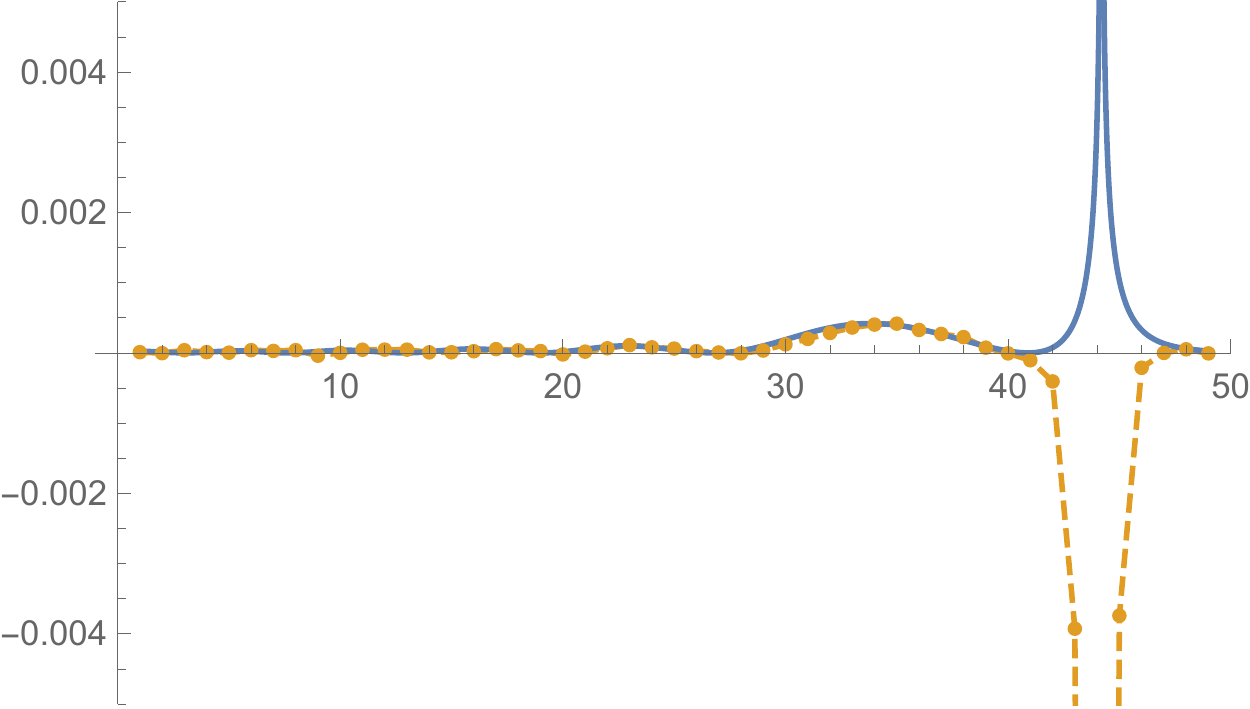}
\caption{$\e1 = 0, \e2 = 0, \widetilde{\e1}=0, \widetilde{\e2}=0$}
\end{subfigure}
\begin{subfigure}[t]{0.45\textwidth}
\centering
\includegraphics[width=\textwidth]{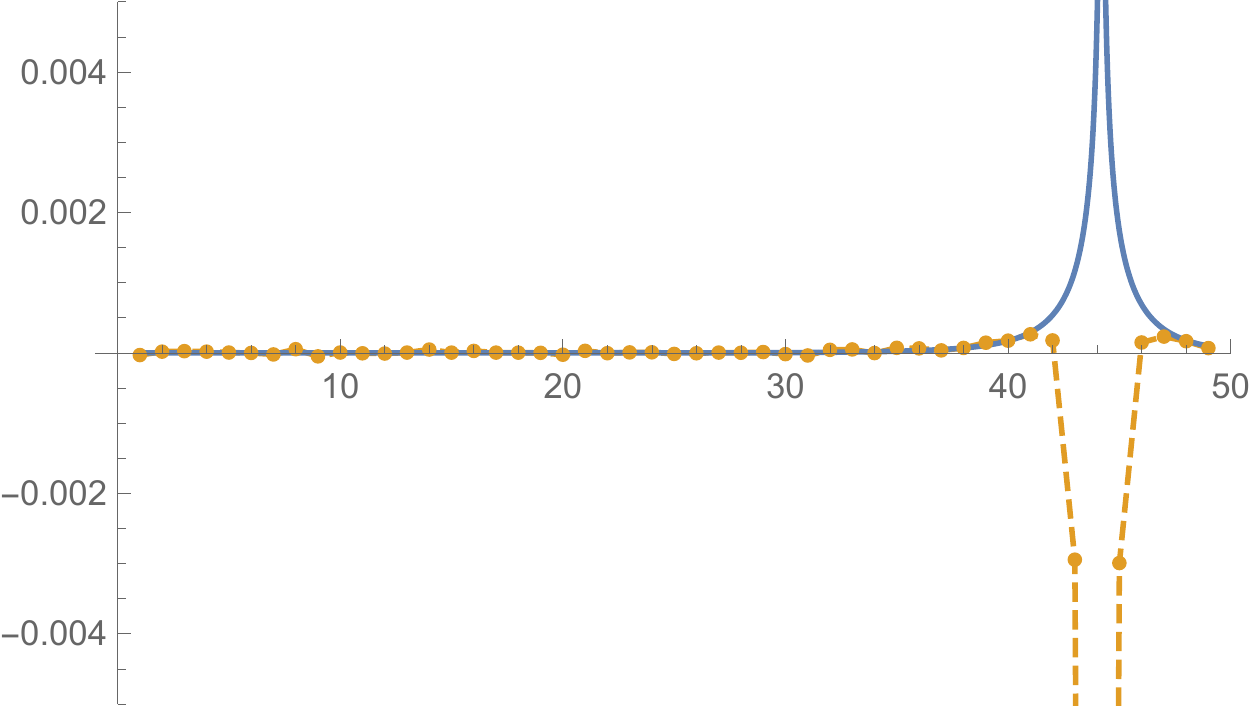}
\caption{$\e1 = 1, \e2 = 1, \widetilde{\e1}=1, \widetilde{\e2}=1$}
\end{subfigure}
\caption{$\widetilde{\alpha} = -0.6$. The blue solid lines show $-m^{-1}s(x,y,\widetilde{x},\widetilde{y}) q_{\e1 \widetilde{\e2}}(\alpha,\widetilde{\alpha}) q_{\widetilde{\e1} \e2}(\widetilde{\alpha},\alpha)$. The orange markers connected by dashed lines show the experimental covariances $\cov(e,\widetilde{e})$, for edges $e,\widetilde{e}$ as shown in Figure~\ref{fig:dimerpairs}. Here, $n=256$ and $a=0.875$.}\label{fig:expalpha06}
\end{figure}

Let \[s(x,y,\widetilde{x},\widetilde{y}) = \frac{\Sigma(x, y) \Sigma(\widetilde{x},\widetilde{y})}{\Sigma(x,\widetilde{y})\Sigma(\widetilde{x},y)}  \zeta(x,\widetilde{y}) \zeta(\widetilde{x},y).\] This takes values in $\{-1,1\}$. For our simulations we take $B = 1$, and plot the experimental covariances against the theorized asymptotic covariances. We fix $\widetilde{\alpha}$ and take $\alpha$ increasing from $-6$ to $0$. We use a size of $n = 256$, and present results for $\widetilde{\alpha} = -3$ and  $\widetilde{\alpha} = -0.6$. We present results for the six pair of types of dimers shown in Figure~\ref{fig:dimerpairs}. Plots of the experimental covariances against $-m^{-1}s(x,y,\widetilde{x},\widetilde{y}) q_{\e1 \widetilde{\e2}}(\alpha,\widetilde{\alpha}) q_{\widetilde{\e1} \e2}(\widetilde{\alpha},\alpha)$ are shown for $\widetilde{\alpha} = -3$ in Figure~\ref{fig:expalpha3} and for $\widetilde{\alpha} = -0.6$ in Figure~\ref{fig:expalpha06}. We see that the conjectured leading order covariances agree well with experiment when $\alpha$ is sufficiently far from $\widetilde{\alpha}$. As discussed in Section~\ref{sec:numericalK} and illustrated in Figures~\ref{fig:k01numeric}--\ref{fig:k11numeric}, we would not expect very good agreement when $\alpha$ is close to $\widetilde{\alpha}$ for tilings of this size.

\backmatter

\bmhead{Acknowledgments}

This research was partly supported by NSF grant DMS-1902226. We are grateful to Nicolai Reshetikhin for guidance during the course of this work. We used code based on the source code developed by Keating and Sridhar \cite{keating2018code} to simulate domino tilings. We thank Scott Mason for correspondence which resulted in a simplification of the paper. We are grateful to the Yau Mathematical Sciences Center, Tsinghua University where much of this work was completed.

This research used the Savio computational cluster resource provided by the Berkeley Research Computing program at the University of California, Berkeley (supported by the UC Berkeley Chancellor, Vice Chancellor for Research, and Chief Information Officer). 

\section*{Declarations}

\subsection*{Competing interests}

The authors have no relevant financial or non-financial interests to disclose.

\subsection*{Availability of data and materials}

Data sharing not applicable to this article as no datasets were generated or analysed during the current study.

\begin{appendices}

\section{Definition of $\mathcal{I}^{j,k}_{\e1,\e2} (a, x_1, x_2, y_1, y_2)$}\label{sec:mathcalI}

 Let
\[c = \frac{1}{a + a^{-1}}.\] For $\omega \in \mathbb{C}\setminus i[-\sqrt{2c},\sqrt{2c}]$ we define \begin{equation*}
\sqrt{\omega^2 + 2c} = i\sqrt{-i(\omega + i\sqrt{2c})}\sqrt{-i(\omega - i \sqrt{2c})}\end{equation*} where the square roots on the right hand side are the principal branch of the square root. Define \begin{equation*}
G(\omega) = \frac{1}{\sqrt{2c}}(\omega - \sqrt{\omega^2 + 2c}).\end{equation*} For even $x_1, x_2$ with $0 < x_1,x_2 < 2n$ define \begin{equation*}
\widetilde{H}_{x_1,x_2}(\omega) = \frac{\omega^{2m}(-iG(\omega))^{2m - x_1/2}}{(iG(\omega^{-1}))^{2m-x_2/2}}.\end{equation*}  For $j,k,\e1,\e2 \in \{0,1\}$, define
\begin{equation*}
V^{j,k}_{\e1, \e2} (\w1, \w2) = \frac{1}{2}\sum_{\gamma_1,\gamma_2=0}^1 (-1)^{\gamma_2 j + \gamma_1 k}(Q_{\gamma_1,\gamma_2}^{\e1,\e2}(\omega_1,\omega_2) + (-1)^{\e2+1}Q_{\gamma_1,\gamma_2}^{\e1,\e2}(\omega_1,-\omega_2))\end{equation*} where the functions $Q_{\gamma_1,\gamma_2}^{\e1,\e2}(\omega_1,\omega_2)$ are defined as follows. Let \begin{equation*}
f_{a,b}(u, v) = (2a^2 uv + 2b^2 uv -ab(-1+u^2)(-1+v^2))
 (2a^2 uv + 2b^2 uv +ab(-1+u^2)(-1+v^2)). 
\end{equation*} Now we define the following rational functions. We temporarily consider weights $a$ and $b$ where $b$ is not necessarily 1. Let
\begin{align*}
\begin{split}
\mathrm{y}_{0,0}^{0,0}(a,b,u,v) =& \frac{1}{4(a^2+b^2)^2 f_{a,b}(u, v)}(2a^7 u^2 v^2 - a^5 b^2(1 + u^4 + u^2v^2 - u^4v^2 + v^4 - u^2v^4) \\
&- a^3 b^4(1+3u^2 + 3v^2 + 2u^2v^2 + u^4v^2 + u^2v^4 - u^4v^4) \\
&- a b^6(1+v^2+u^2+3u^2v^2)) \\
\mathrm{y}_{0,1}^{0,0}(a,b,u,v) =& \frac{a}{4(a^2 + b^2)f_{a,b}(u, v)}(b^2 + a^2u^2)(2a^2v^2 + b^2(1 + v^2 - u^2 + u^2v^2)) \\
\mathrm{y}_{1,0}^{0,0}(a,b,u,v) =& \frac{a}{4(a^2+b^2)f_{a,b}(u, v)}(b^2 + a^2v^2)(2a^2u^2 + b^2(1 - u^2 + v^2 + u^2v^2)) \\
\mathrm{y}_{1,1}^{0,0}(a,b,u,v) =& \frac{a}{4f_{a,b}(u, v)}(2a^2u^2v^2 +b^2(- 1 + v^2 + u^2 + u^2v^2)).\end{split}
\end{align*} For $\gamma_1,\gamma_2 \in \{0,1\}$ we define  \begin{align*}
\begin{split}
\mathrm{y}_{\gamma_1,\gamma_2}^{0,1}(a,b,u,v) &= \frac{\mathrm{y}_{\gamma_1,\gamma_2}^{0,0}(b,a, u,v^{-1})}{v^2} \\
\mathrm{y}_{\gamma_1,\gamma_2}^{1,0}(a,b,u,v) &= \frac{\mathrm{y}_{\gamma_1,\gamma_2}^{0,0}(b,a,u^{-1},v)}{u^2} \\
\mathrm{y}_{\gamma_1,\gamma_2}^{1,1}(a,b,u,v) &= \frac{\mathrm{y}_{\gamma_1,\gamma_2}^{0,0}(a,b,u^{-1},v^{-1})}{v^2}.
\end{split}
\end{align*} When $b=1$, we write $\mathrm{y}^{\e1,\e2}_{\gamma_1,\gamma_2}(u,v)  = \mathrm{y}^{\e1,\e2}_{\gamma_1,\gamma_2}(a,1,u,v) $. Then define \begin{equation*}\mathrm{x}_{\g1, \g2}^{\e1, \e2}(\w1, \w2) = \frac{G(\w1) G(\w2)}{\prod_{i=1}^2 \sqrt{\omega_i^2 +2c} \sqrt{\omega_i^{-2} + 2c}} \textrm{y}_{\g1, \g2}^{\e1, \e2} (G(\w1), G(\w2))(1 - \w1^2\w2^2). \end{equation*} and \begin{multline*}
Q_{\g1,\g2}^{\e1,\e2}(\w1,\w2) = (-1)^{\e1+\e2+\e1\e2 + \g1(1+\e2) + \g2(1+\e1)}\\
\times t(\w1)^{\g1}t(\w2^{-1})^{\g2} G(\w1)^\e1 G(\w2^{-1})^\e2 \mathrm{x}_{\g1, \g2}^{\e1, \e2}(\w1, \w2^{-1})\end{multline*} where $t(\omega)$ is defined by \begin{equation*}
t(\omega) = \omega\sqrt{\omega^{-2} + 2c}.\end{equation*}
 For  $x = (x_1,x_2) \in \mathtt{W}_\e1,$ and $y = (y_1,y_2) \in \mathtt{B}_\e2$ with $\e1,\e2\in \{0,1\}$, define \begin{align*}
\begin{split}
h_{0,0}(\w1,\w2) &=  \frac{\widetilde{H}_{x_1 + 1, x_2}(\w1)}{\widetilde{H}_{y_1, y_2 + 1}(\w2)}\\
h_{1,0}(\w1,\w2) &= \frac{\widetilde{H}_{x_1 + 1, x_2}(\w1)}{\widetilde{H}_{2n-y_1, y_2 + 1}(\w2)} \\
h_{0,1}(\w1,\w2) &= \frac{\widetilde{H}_{x_1 + 1, 2n - x_2}(\w1)}{\widetilde{H}_{y_1, y_2 + 1}(\w2)} \\
h_{1,1}(\w1,\w2) &=  \frac{\widetilde{H}_{x_1 + 1, 2n - x_2}(\w1)}{\widetilde{H}_{2n-y_1, y_2 + 1}(\w2)}
\end{split}
\end{align*}
Let $\circlecontour_r$ denote a positively-oriented contour of radius $r$ centered at the origin. For $a < 1$, $\sqrt{2c} < r < 1$ and $x = (x_1,x_2) \in \mathtt{W}_\e1,\, y = (y_1,y_2) \in \mathtt{B}_\e2$ with $\e1,\e2\in \{0,1\}$ define \begin{align}\label{eq:mathcalI}
\mathcal{I}^{j,k}_{\e1,\e2} (a, x_1, x_2, y_1, y_2) &= \frac{i^{y_1-x_1}}{(2\pi i)^2} \int_{\circlecontour_r} \frac{d\w1}{\w1} \int_{\circlecontour_{1/r}}d\w2 \frac{V_{\e1, \e2}^{j,k} (\w1, \w2)}{\w2 - \w1} h_{j,k}(\w1,\w2).
\end{align}

\end{appendices}


\bibliography{sn-bibliography}

\end{document}